\documentclass[fleqn,usenatbib]{mnras}

\usepackage{newtxtext,newtxmath}

\usepackage[T1]{fontenc}

\usepackage{amsmath}

\usepackage{tabularx}
\usepackage{multirow}
\usepackage{siunitx}
\usepackage{upgreek}

\usepackage{graphicx} 
\graphicspath{{figures/}}
\usepackage{easyReview}

\usepackage[noabbrev,nameinlink]{cleveref}
\Crefname{figure}{Fig.}{Figs.}

\title[Minkowski Functionals of CMB polarized intensity]{Minkowski Functionals of CMB polarization intensity with Pynkowski: theory and application to \textit{Planck} and future data}
\author[A. Carones et al.]{
Alessandro Carones$^{1,2}$\thanks{E-mail: alessandro.carones@roma2.infn.it},
Javier Carrón Duque$^{1,2}$,
Domenico Marinucci$^{3}$,
Marina Migliaccio$^{1,2}$,
Nicola Vittorio$^{1,2}$
\\
$^{1}$Dipartimento di Fisica, Universit\`a di Roma ``Tor~Vergata'', via della Ricerca Scientifica 1, I-00133, Roma, Italy \\
$^{2}$Sezione INFN Roma~2, via della Ricerca Scientifica 1, I-00133, Roma, Italy \\
$^{3}$Dipartimento di Matematica, Universita’ di Roma Tor Vergata, via della Ricerca Scientifica 1, I-00133, Roma, Italy
}

\date{Accepted XXX. Received YYY; in original form ZZZ}

\pubyear{2022}

\begin{document}
\label{firstpage}
\pagerange{\pageref{firstpage}--\pageref{lastpage}}
\maketitle

\begin{abstract}
The angular power spectrum of the Cosmic Microwave Background (CMB) anisotropies is a key tool to study the Universe. However, it is blind to the presence of non--Gaussianities and deviations from statistical isotropy, which can be detected with other statistics such as Minkowski Functionals (MFs). These tools have been applied to CMB temperature and $E$-mode anisotropies with no detection of deviations from Gaussianity and isotropy. In this work, we extend the MFs formalism to the CMB polarization intensity, $P^2=Q^2+U^2$. We use the Gaussian Kinematic Formula to derive the theoretical predictions of MFs for Gaussian isotropic fields. We develop a software that computes MFs on $P^2$ HEALPix maps and apply it to simulations to verify the robustness of both theory and methodology. We then estimate MFs of $P^2$ maps from \textit{Planck}, both in pixel space and needlet domain, comparing them with realistic simulations which include CMB and instrumental noise residuals. We find no significant deviations from Gaussianity or isotropy in \textit{Planck} CMB polarization intensity. However, MFs could play an important role in the analysis of CMB polarization measurements from upcoming experiments with improved sensitivity. Therefore we forecast the ability of MFs applied to $P^2$ maps to detect much fainter non-Gaussian anisotropic signals than with \textit{Planck} data for two future complementary experiments: the \textit{LiteBIRD} satellite and the ground-based Simons Observatory. We publicly release the software to compute MFs in arbitrary scalar HEALPix maps as a fully-documented Python package called \href{https://github.com/javicarron/pynkowski}{\texttt{Pynkowski}}.

	\end{abstract}
	

\section{Introduction}
\label{s:intro}

A wide variety of powerful statistical and geometrical tools have been adopted for the analysis of Cosmological data. Among them, the angular power spectrum of the anisotropies of the Cosmic Microwave Background (CMB), both in temperature and polarization, has played a key role in constraining the values of the parameters of the $\Lambda$CDM cosmological model \citep{2020A&A...641A...6P}. The angular power spectrum contains all the statistical information of a Gaussian isotropic field, but it is blind to the presence of non-Gaussianities or physical anisotropies in the data; in other words, two observed maps can have the same angular power spectrum but still have widely different statistical properties. 

Searching for the presence of small non--Gaussianities or deviations from statistical isotropy in the primordial Universe is an active field of research, as they could encode information about Cosmic Inflation and physics of the very Early Universe, if detected \citep{bartolo2004, pitrou2008, barrow2006, 2010AdAst2010E..72C}. Such effects, however, are not associated only to the physics of the primordial Universe: CMB maps are contaminated by residual foregrounds, complex instrumental systematics, and non-primordial cosmological effects such as CMB lensing, which can induce significant deviations from the Gaussianity and isotropy hypotheses. These signals contain relevant information which is not reflected in the angular power spectrum. Non--Gaussian statistics of CMB lensing have already been shown to help constrain cosmological parameters \citep{zuercher2021, 2020ApJ...897...14C, 2022OJAp....5E..13G}. 

Recently, many studies about primordial non-Gaussianity have also involved the analysis of the Large Scale Structure of the Universe \citep{2022arXiv220308838A, 2022MNRAS.511.2259P}. In particular, the high-order statistics of galaxy clustering with future galaxy surveys could improve the constraints on primordial non--Gaussianity \citep{2010CQGra..27l4011D}.
     
Minkowski Functionals (MFs) are one of the most popular tools used to analyse information complementary to that contained in the angular power spectrum. MFs have been widely explored in the mathematical literature \citep{tomita1986,coles1987,gott1990}, and have become a popular tool in CMB analysis after the discussion by \citet{schmalzing1998}. An extension of the MFs formalism to study the full information embedded in the spin--$2$ CMB polarization field has recently been proposed in \citet{2023carron}.

In the case of scalar Gaussian isotropic fields such as the CMB temperature anisotropies, the theoretical predictions of these functionals can be accurately computed. Given their low variance, any deviation from Gaussianity or statistical isotropy can be detected at high significance in a model--independent manner. MFs are now widely used to measure possible deviations from primordial Gaussianity \citep{2020A&A...641A...7P} and, beyond that, \textit{e.g.}, to constrain cosmological parameters \citep{zuercher2021}. Moreover, these  tools can be applied to detect contamination by foreground and instrumental systematics in the reconstructed CMB map in a fully blind fashion, given their highly non-Gaussian and anisotropic nature. The impact of such systematic and astrophysical effects on the constraining power for primordial non-Gaussianity in temperature data has also been explored by \citet{ducout2012}. Up to now, MFs have been largely applied to the scalar fields associated to the CMB polarization, E-modes and B-modes, \citep{2016A&A...594A..16P,2015JCAP...02..028G,2016JCAP...07..029S,2020A&A...641A...7P}, while they remain mostly unexploited in the analysis of spin fields. Therefore, in this work we aim to extend the formalism of MFs to the squared modulus of the CMB polarization, $P^2=Q^2+U^2$, which hereafter we will refer to simply as polarization modulus or polarized intensity. 

This new formalism conveys complementary information on the CMB polarization field with respect to the study of E- and B-modes in several aspects:
\begin{itemize}
    \item The polarization modulus traces the local properties of the polarization field, while the E- and B-modes maps are obtained from a non-local decomposition of Q and U, typically performed in harmonic space.
    \item The $P^2$ field can be robustly obtained for any adopted mask without suffering of E--B and B--E leakage contamination \citep{2001PhRvD..65b3505L,2003PhRvD..67b3501B}, which may distort the statistics of the reconstructed E- and B-modes.
    \item The morphology of the polarization modulus is intrinsically different and can be more effective in detecting those deviations from Gaussianity and statistical isotropy which primarily affect the intensity rather than the direction of the polarized emission.
    
\end{itemize}

We note that, although E- and B-mode maps embed the full information on the polarization field (in the full-sky case), analysing MFs of these fields independently do not exhaust the morphological characterisation of the polarization field. This motivates the usefulness of analysing complementary quantities as $P^2$.

This paper has $5$ key goals:
\begin{enumerate}
    \item to introduce a more general formalism to compute the theoretical predictions of scalar maps, such as the CMB temperature anisotropies ($T$),
    \item to apply this extended formalism to obtain the expected values of the MFs of the modulus of a spin field, such as the CMB polarization ($P^2$) in the case of Gaussianity and statistical isotropy,
    \item to provide the community with a public Python package called \texttt{Pynkowski}\footnote{\url{https://github.com/javicarron/pynkowski}} to easily estimate MFs on HEALPix\footnote{\url{https://github.com/healpy/healpy}} maps \citep{2005ApJ...622..759G} and compute their Gaussian isotropic expectations.
    \item to compute the MFs on \textit{Planck} $P^2$ maps and compare them with those from realistic simulations in order to assess any possible unexpected deviation from Gaussianity or isotropy.
    \item to forecast the sensitivity of MFs applied to $P^2$ maps to deviations from Gaussianity and statistical isotropy for future CMB experiments.
\end{enumerate}
We note that, up to second--order terms, theoretical predictions of MFs for the sum of two squared Gaussian fields have already been presented in the literature
    \citep[see][and the references therein]{naselsky1998}. However, in this work these results are further justified and are derived by means of a more general approach. Furthermore, for the first time these predictions are compared with realistic CMB simulations, and the MFs of the $P^2$ field are estimated on \textit{Planck} data. These CMB polarization maps are dominated by the anisotropic noise, however we expect these tools to grow in importance for the analysis of data with much greater sensitivity from ongoing or future suborbital and satellite experiments such as ACT \citep{2020JCAP...12..047A}, SPT \citep{2020PhRvD.101l2003S}, LSPE \citep{2021JCAP...08..008A}, Simons Observatory \citep{2019JCAP...02..056A}, CMB-S4 \citep{2022arXiv220308024A},
and \textit{LiteBIRD} \citep{2014JLTP..176..733M}. To further investigate this point, we forecast the capability of MFs computed on simulated realistic $P^2$ maps to detect deviations from Gaussianity and statistical isotropy for two future and complementary CMB polarization experiments: Simons Observatory (SO) and \textit{LiteBIRD}. Both of them will target the detection of the so-called primordial CMB polarization $B$ modes, supposed to be generated by a stochastic background of gravitational waves from cosmic inflation. SO is ground-based and located in the Atacama Desert in Chile. Due to its limited observed sky fraction ($f_{\textrm{sky}} \sim 30 \%$) and the atmospheric contamination, SO is sensitive to polarization modes with $\ell > 30$ and thus will only target the detection of the recombination peak in the primordial $B$-mode power spectrum at $\ell \sim 100$ \citep{2019JCAP...02..056A}. \textit{LiteBIRD}, instead, is a space-borne L-class mission selected by the Japanaese Aerospace Exploration Agency (JAXA) in $2019$. Given its full-sky coverage, it will be sensitive also to modes on the largest angular scales \citep{PTEPlitebird}. For both experiments, the detection of primordial CMB $B$ modes will be hampered by residual contamination of the Galactic emission, which is highly non-Gaussian and anisotropic. Therefore we have included such a signal in the \textit{LiteBIRD} and SO-SATs polarization simulated data as non-Gaussian anisotropic component.

In this paper, we also present, to our knowledge, the first public software needed to compute these MFs on any HEALPix map. The structure of this article is as follows. In \Cref{s:mf_T} we present the definition of MFs and their application to stochastic fields defined on the sphere. In \Cref{s:mf_P} we derive the theoretical expectations of the MFs for CMB polarized intensity: $P^2 = Q^2 + U^2$. In \Cref{s:code} we present the \texttt{Pynkowski} Python package and explain the procedures implemented to estimate these statistical quantities on the maps. In \Cref{s:data} we introduce the data and simulations we use in this paper. In \Cref{s:results} we present the results of applying this framework and software to CMB polarization simulations and \textit{Planck} observations, as well as the forecasted sensitivity of $P^2$ MFs to a non-Gaussian anisotropic signal for \textit{LiteBIRD} and SO. Finally, in \Cref{s:concl} we report our conclusions.
    	
\section{Minkowski Functionals on the Sphere}
\label{s:mf_T}

MFs on the sphere are now well--established tools for CMB data analysis \citep{2020A&A...641A...7P}; nonetheless, we will first recall some background notation and definitions in order to make the comparison with our present work more explicit. Given a scalar random field $f(.)$ observed on the unit sphere $\mathbb{S}^{2}$ (\textit{e.g.}, CMB temperature anisotropies), the excursion set at a threshold $u$ is defined by%
\[
A_{u}(f(.),\mathbb{S}^{2})=\left\{ x\in \mathbb{S}^{2} : f(x)\geq u\right\}
\text{ .}
\]%
We will omit the arguments of $A_u$ to refer to the excursion sets when $f$ and the domain of $f$ are clear by context.

As recalled in \citet{schmalzing1998}, the morphological properties of these excursion sets can be summarised by the three MFs defined as follows:
\begin{align*}
V_{0}(A_{u}) &=\int_{A_{u}}dx\text{ , }\\
V_{1}(A_{u}) &= \frac{1}{4}%
\int_{\partial A_{u}}dr\text{ , } \\
V_{2}(A_{u}) &=\frac{1}{2\pi }\int_{\partial A_{u}}\kappa (r)dr\text{ ,}
\end{align*}%
where $dr$ denotes a line element along the boundary of the excursion set and $\kappa (.)$ denotes the geodesic curvature of said boundary. Thus, $V_{0}$ represents the excursion area, while $V_{1}$ is one fourth of the boundary length. $V_{2}$ is associated with the number of connected regions minus the number of holes, which on the plane corresponds to the Euler--Poincar\'e characteristic $\chi$. On a spherical surface the Euler--Poincar\'e characteristic does no longer correspond exactly to $V_2$, but includes an extra term which can be computed
by means of the generalised Gauss--Bonnet Theorem, which yields:%
\[
\chi (A_{u})=V_{2}(A_{u})+\frac{1}{2\pi }V_{0}(A_{u})\text{ .}
\]%
From the mathematical point of view, it turns out to be notationally more
convenient to replace MFs with the equivalent notion of
Lipschitz--Killing Curvatures, which are defined by%
\begin{align}
\mathcal{L}_{2}(A_{u}) &=V_{0}(A_{u})\text{ ,} \\
\mathcal{L}_{1}(A_{u}) &=2V_{1}(A_{u})\text{ ,} \\
\mathcal{L}_{0}(A_{u}) &=\chi (A_{u})\text{ .}
\label{eq:LK}
\end{align}%

These quantities are defined as follows. Let us define a tube of width $\rho$ built around the manifold $A_{u}$ as all those points at a distance less than $\rho$ from $A_{u}$. Its volume can be exactly expressed as a finite Taylor expansion on $\rho$, whose coefficients correspond to the Lipschitz--Killing curvatures of $A_{u}$ (see \citealt{adler2007,fantaye2014} for a detailed discussion). 

Let us now recall how to compute the expected values of these statistical quantities for Gaussian isotropic maps. These results are well--known in the Cosmological literature in the case of the two-dimensional sphere, but it is convenient to recall them from a more general perspective. 

Without loss of generality we assume the field normalised to have unit variance. We define $\mu $ as the derivative of the covariance function at the origin for the field $f$. This quantity can be computed as follows:%
\[
\mu =\sum_{\ell }\frac{\ell (\ell +1)}{2}\frac{2\ell +1}{4\pi }C_{\ell }%
\text{ , }
\]%
where the sequence $\left\{ C_{\ell }\right\} $ denotes, as usual, the angular power spectrum of the field $f$. Furthermore, following \citet{adler2007} we define the \textit{flag} coefficients as%
\[
\left[
\begin{array}{c}
k+j \\
k%
\end{array}%
\right] =\frac{\omega _{k+j}}{\omega _{k}\omega _{j}}\left(
\begin{array}{c}
k+j \\
k%
\end{array}%
\right) \text{ , } \qquad 
\omega _{j}=\frac{\pi ^{j/2}}{\Gamma (\frac{j}{2}+1)}\text{
,}
\]%
with $\omega _{1}=2$, $\omega _{2}=\pi$, $\omega _{3}=\frac{4\pi }{3}$. In general, $\omega_j$
represents the volume of the $j$-dimensional unit ball. For an isotropic Gaussian field, the expected value of the Lipschitz-Killing Curvatures on the sphere is then given by the Gaussian Kinematic Formula (GKF):%
\begin{equation}
\mathbb{E}\left[ \mathcal{L}_{j}(A_{u})\right]
=\sum_{k=0}^{2-j}\left[
\begin{array}{c}
k+j \\
k%
\end{array}%
\right] \rho _{k}(u)\mathcal{L}_{k+j}(\mathbb{S}^{2})\mu ^{k/2},  \label{GKF}
\end{equation}%
where%
\begin{align*}
\rho _{k}(u) =\frac{1}{(2\pi )^{k/2}}&\frac{1}{\sqrt{2\pi }}\exp\left(-\frac{%
u^{2}}{2}\right)H_{k-1}(u)\text{ ,} \\
H_{-1}(u) &=\sqrt{2\pi}\exp\left(\frac{%
u^{2}}{2}\right)\left(1-\Phi (u)\right)\text{ , }\\H_{0}(u)&=1\text{ , }\\H_{1}(u)&=u\text{ ,}
\end{align*}%
and in general, $H_{k}(u)=(-1)^{k}\exp \Big(\frac{u^{2}}{2}\Big)\frac{d^{k}}{du^{k}}\exp \Big(-%
\frac{u^{2}}{2}\Big)$ denotes the sequence of Hermite polynomials. The function $\Phi$ is the cumulative normal distribution.

In particular, by \cref{GKF} we get the well--known results for the expected values of the Lipschitz--Killing Curvatures and, using \cref{eq:LK}, of MFs, which we report normalised by area. In order, we have:%
\[
\mathbb{E}\left[ \mathcal{L}_{2}(A_{u})\right] =\rho
_{0}(u)\mathcal{L}_{2}(\mathbb{S}^{2})=4\pi \left\{ 1-\Phi (u)\right\} \text{
,}
\]%
\[
\frac{\mathbb{E}\left[V_{0}\right]}{4\pi }=1-\Phi (u)\text{ .}
\]
We note that this quantity depends only on $u$, not on the field itself as long as it is normalised to have unit variance. 

Likewise:%
\begin{align*}
\mathbb{E}\left[ \mathcal{L}_{1}(A_{u})\right]  &=2\frac{\omega _{2}}{%
\omega _{1}^{2}}\rho _{1}(u)\mathcal{L}_{2}(\mathbb{S}^{2})\mu ^{1/2} \\
&=\frac{2\pi }{4}\frac{1}{(2\pi )^{1/2}}\frac{1}{\sqrt{2\pi }}\exp \Bigg(-\frac{%
u^{2}}{2}\Bigg)4\pi \mu ^{1/2} \\
&=\exp \Bigg(-\frac{u^{2}}{2}\Bigg)\pi \mu ^{1/2}
\end{align*}%
so that%
\[
\frac{\mathbb{E}\left[V_{1}\right]}{4\pi }=\frac{1}{8}\exp \Bigg(-\frac{u^{2}}{2}\Bigg)\mu ^{1/2}%
\text{ .}
\]%
Finally:%
\begin{align*}
\mathbb{E}\left[ \mathcal{L}_{0}(A_{u})\right]  &=\rho _{0}(u)\mathcal{L}%
_{0}(\mathbb{S}^{2})+\rho _{2}(u)\mathcal{L}_{2}(\mathbb{S}^{2})\mu  \\
&=\left\{ 1-\Phi (u)\right\} 2+\frac{2}{\sqrt{2\pi }}\mu \exp \Bigg(-\frac{u^{2}}{2}%
\Bigg)u\text{ ,}
\end{align*}%
and hence:%
\[
\frac{\mathbb{E}\left[V_{2}\right]}{4\pi }=
\frac{1}{\sqrt{(2\pi )^{3}}} \mu \exp \Bigg(-\frac{u^{2}}{2}\Bigg)u\text{ .}
\]%

The interpretation of the previous formulae is worth discussing: the expected values of the MFs (and Lipschitz--Killing Curvatures) are expressed by means of a product 
of four different independent ingredients:

\begin{enumerate}
    \item a set of universal coefficients (called ``flag'' coefficients),
    \item the Lipschitz--Killing Curvatures evaluated on the original manifold (in our case, the unitary sphere),
    \item a power of the derivative of the covariance function of the field at the origin,
    \item a set of ``density'' functions $\rho_k$, dependent only on the threshold at which the MFs (or Lipschitz--Killing Curvatures) are evaluated.
\end{enumerate}

The remarkable fact is that in the case of polarization, the expected values for Lipschitz--Killing Curvatures take exactly the same form---the only change is in the ``density'' functions $\rho_k$, which nonetheless can be expressed with a fully explicit analytic form. We give more details in the section to follow. 

We also note that the presence of a mask only affects the manifold where the map is defined, and therefore, at leading order, it has an impact only on the normalisation of the MFs.

\section{Minkowski Functionals on polarization Fields}
\label{s:mf_P}

A spin $s$ random field $P_s$ is a complex--valued field that satisfies the following relation at any fixed point $x$: %
\[
P_{s}^{\prime }(x)=\exp (is\theta )P_s(x)\text{ .}
\]%
where $P^{\prime }(x)$ represents the value in $x\in \mathbb{S}^{2}$ when the local coordinates are rotated by an angle $\theta$.
This notion can be made mathematically rigorous viewing spin fields as a section of a so--called spin fibre bundle, first introduced in mathematical physics by \citet{newman1966}.

Under this definition, the CMB polarization is a Gaussian
complex--valued spin-2 random field, that can be written as:%
\[
P(x)=Q(x)+iU(x)\text{ , }P^{\prime }(x)=\exp (i2\theta )P(x)\text{ ,}
\]%
where $Q$ and $U$ are real maps. 
It is important to note that, for any fixed choice of a coordinate system, polarization fields can be viewed as complex--valued Gaussian random variables at a single point, but they can no longer be considered as the realisation of an isotropic map. Indeed the correlation function at any two points is not invariant to rotations, and hence it does not only depend on the geodesic distance between the two points (as required by isotropy). Only in the
special case where $s=0$, we are back to the case of a complex--valued isotropic Gaussian map, which can always be viewed as the complexification of two Gaussian independent and identically distributed real--valued fields $Q$ and $U$. 

The behaviour of the expected values for Lipschitz--Killing Curvatures and other topological functionals of spin fields has been very recently investigated in the mathematical literature by \citet{lerario2022}. The main finding of these authors is that for ``band--limited'' spin--valued random fields with power spectrum concentrated around multipoles equal or larger than $\ell_*$, the effect of the spin vanishes as $\ell_{*}$ grows. In that case, these geometric functionals have the same behaviour for spin Gaussian fields and for complex--valued scalar Gaussian fields. More precisely, the remainder terms (neglected when approximating the geometry of a spin $s \neq 0$ field with the spin $s=0$ case) are of the form $\propto\frac{s}{\ell(\ell+1)}$. Thus, spin effects can be safely ignored for the random fields of cosmological interest, as there is very little power at small multipoles. Incidentally, from the mathematical point of view, these results continue to hold for any fixed value of the spin parameter $s$, and even under other conditions such as the (non--physical) case of a spin parameter growing with $\ell_{*}$.
We can informally summarise this mathematical result as follows:

\medskip
\emph{Up to negligible terms which depend only on the smallest multipoles, the expected values of the Lipschitz--Killing Curvatures for the excursion sets of the norms of spin }$s$\emph{\ random
fields are identical, for all values of }$s.$\emph{\ In other words,
defining }%
\[
|P_{s}(x)|^{2}:=Q_{s}^{2}(x)+U_{s}^{2}(x)\text{ ,}
\]%
\emph{we have that}%
\[
\mathbb{E}\left[ \mathcal{L}_{j}(A_{u}(|P_{s}(.)|^{2},\mathbb{S}^{2})\right]
=\mathbb{E}\left[ \mathcal{L}_{j}(A_{u}(|P_{0}(.)|^{2},\mathbb{S}^{2})\right]
\text{ ,}
\]%
\emph{for all values of }$j$\emph{\ and }$s.$ 
\\
\medskip

We are now going to derive the expected values of Lipschitz--Killing Curvatures using the GKF, in an analogous way to the previous Section. It should be noted that, up to second--order terms, results equivalent to ours have been given before in the literature, see \citet{kasak2020}, \citet{naselsky1998}, and the references therein. Apart from putting these results into more solid grounds by means of the approximation that we just explained, we believe that our derivation by means of the GKF is clearer and more amenable to generalisations.

In particular, assume that we have $Q$ and $U$ two independent copies of Gaussian random fields with the same angular power spectrum, expected value $0$ and variance normalised to 
$1$ (\textit{i.e.} $\mathbb{E}[Q^{2}(x)]=\mathbb{E}[U^{2}(x)]=1$). Defining%
\[
P^{2}(x)=Q^{2}(x)+U^{2}(x)\text{ ,}
\]%
it is immediate to demonstrate that $P^{2}$ is a chi-squared with two degrees of freedom, $P^{2}\sim \chi _{2}^{2},$ and hence it has marginal density%
\[
f_{P^{2}}(u)=\frac{1}{2}\exp \Big(-\frac{u}{2}\Big)\text{ .}
\]%
We can now consider the excursion sets
\[
A_{u}(P^{2},\mathbb{S}^{2})=\left\{ x\in \mathbb{S}^{2}:P^{2}(x)\geq u\right\}
\text{ .}
\]%

We can now recall again the GKF, which in the case of chi-square fields reads%
\begin{align}
\begin{split}
    \mathbb{E}[& \mathcal{L}_{j}(A_{u}(P^{2},\mathbb{S}^{2}))]
    =\\
    &=\sum_{k=0}^{d-j}\left[
    \begin{array}{c}
    k+j \\
    k%
    \end{array}%
    \right] \rho _{k;P^2}(u)\mathcal{L}_{k+j}(\mathbb{S}^{2})\mu ^{k/2}\text{ ,}
\end{split}
\label{GKF2}
\end{align}
where as usual we write $\mu $ for the derivative of the covariance function evaluated at the origin. 
It should be noted that the expression for the expected values takes exactly the same form as in \cref{GKF}, simply replacing the functions $\rho _{k}(.)$ with $\rho
_{k;P^2}(.)$ defined by \citet[][Theorem 15.10.1]{adler2007}:%
\begin{align*}
\rho _{j,P^2}(u) &=\frac{u^{(k-j)/2}\exp ^{-u/2}}{(2\pi )^{j/2}\Gamma
(k/2)2^{(k-2)/2}}\\
&\times
\sum_{l=0}^{\left\lfloor \frac{j-1}{2}\right\rfloor
} \sum_{m=0}^{j-1-2l} \mathbb{I}_{\left\{ k\geq j-m-2l\right\} }\binom{k-1}{j-1-m-2l} \\
&\times
\frac{(-1)^{j-1+m+l}}{m!l!2^{l}}u^{m+l}\text{ .}
\end{align*}%
More explicitly, for $j=1$ we obtain%
\[
\rho _{1,P^2}(u)=\frac{u^{1/2}\exp ^{-u/2}}{(2\pi )^{1/2}} \text{ ,}
\]%
whereas for $j=2$ we have
\begin{align*}
\rho _{2,P^2}(u) &=\frac{\exp ^{-u/2}}{(2\pi )}\sum_{m=0}^{1}\binom{1}{1-m}%
(-1)^{1+m}u^{m} \\
&=-\frac{\exp ^{-u/2}}{(2\pi )}+\frac{\exp ^{-u/2}u}{(2\pi )}\\
&=\frac{%
(u-1)\exp (-u/2)}{2\pi }\text{ .}
\end{align*}%
Plugging these expressions into \cref{GKF2} we obtain, for the boundary length,%
\begin{align*}
    \mathbb{E}[ 2\mathcal{L}_{1}&(A_{u}(P^2(.),\mathbb{S}^{2})) ]=\\
    &=2\sum_{k=0}^{d-j}\left[
    \begin{array}{c}
    k+j \\
    k%
    \end{array}%
    \right] \rho _{P^2}(u)\mathcal{L}_{k+j}(\mathbb{S}^{2})\sqrt{\mu } \\
    &=2\left[
    \begin{array}{c}
    2 \\
    1%
    \end{array}%
    \right] \frac{u^{1/2}\exp ^{-u/2}}{(2\pi )^{1/2}}\mathcal{L}_{2}(\mathbb{S}%
    ^{2})\sqrt{\mu } \\
    &=4\frac{\omega _{2}}{\omega _{1}^{2}}\frac{u^{1/2}\exp ^{-u/2}}{(2\pi)^{1/2}}\cdot 4\pi \cdot \sqrt{\mu } \\
    &=4\frac{\pi }{4}\frac{u^{1/2}\exp ^{-u/2}}{(2\pi )^{1/2}}\cdot 4\pi\cdot \sqrt{\mu } \\
    &=\frac{\sqrt{2\pi }}{2}u^{1/2}\exp ^{-u/2}\cdot 4\pi \cdot \sqrt{\mu }%
    \text{ .}
\end{align*}%
The excursion area can be immediately computed as:%
\begin{align*}
\mathbb{E}\left[ \mathcal{L}_{2}(A_{u}(P^2(.),\mathbb{S}^{2}))\right]  &=%
\mathbb{E}\left[ \int_{\mathbb{S}^{2}}\mathbb{I}(P^2(x)\geq u)dx\right]  \\
&=\int_{\mathbb{S}^{2}}\mathbb{E}\left[ \mathbb{I}(P^2(x)\geq u)\right] dx \\
&=4\pi \exp (-u/2)\text{ .}
\end{align*}%
Finally, for the Euler-Poincar\'e Characteristic%
\[
\mathbb{E}\left[ \mathcal{L}_{0}(A_{u}(P^2,\mathbb{S}^{2}))\right] =
\]
\[
=2\mu (u-1)\exp (-u/2) +2\exp \Big(-\frac{u}{2}\Big)\text{ .}
\]

Passing from the Lipschitz--Killing curvatures to MFs and normalising by the area of the sphere, we finally obtain the following predictions for a chi-squared map with $2$ degrees of freedom, such as $P^2$:
\begin{subequations}
\begin{align}
	\frac{\mathbb{E}\left[{V_{0}(A_u)}\right]}{4\pi } & = \exp \Big(-\frac{u}{2}\Big) \\
	\frac{\mathbb{E}\left[{V_{1}(A_u)}\right]}{4\pi } & = \frac{\sqrt{2\pi }}{8}\sqrt{\mu u}\exp \Big(-\frac{u}{2}\Big) \\ 
	\frac{\mathbb{E}\left[{V_{2}(A_u)}\right]}{4\pi } & = \mu \frac{(u-1)\exp (-u/2)}{2\pi }
\end{align}
\label{eqs:theo_exp}
\end{subequations}
where the threshold is non-negative, $u\geq0$. For negative thresholds, it follows from the definition that $V_0$ is exactly $1$, while $V_1$ and $V_2$ are exactly $0$. In particular, $V_2$ presents a discontinuity at $u=0$ connected with the number of non--polarized points \citep[see][]{kasak2020}.

In all these expressions, the variance of the field is unity and $\mu$ is
the derivative of the covariance of the field at the origin, that can be computed as before:  %
\[
\mu =\sum_{\ell }\frac{2\ell +1}{4\pi }\frac{\ell (\ell +1)}{2}C_{\ell }%
\text{ , }\qquad \sum_{\ell }\frac{2\ell +1}{4\pi }C_{\ell }=1\text{ ,}
\]
where:
\[
C_{\ell } = \frac{1}{2}\Big(C^{EE}_{\ell }+C^{BB}_{\ell }\Big),
\]
with $C^{EE}_{\ell }$ and $C^{BB}_{\ell }$, respectively, the EE and BB angular power spectra computed from Q and U maps \citep{1997PhRvL..78.2058K, 1997PhRvD..55.1830Z}.

\section{Pynkowski}
\label{s:code}
We develop a Python package called \texttt{Pynkowski} to compute MFs on spherical maps formatted in the HEALPix convention \citep{gorski2005}. We make this package publicly available to the research community in \url{https://github.com/javicarron/pynkowski}. 

\texttt{Pynkowski} can be used in two different ways:
\begin{enumerate}
    \item To compute the expected values of the MFs for Gaussian scalar fields, like CMB temperature anisotropies $T$ (see \Cref{s:mf_T}), or for $\chi^2$ fields, like CMB polarized intensity $P^2$ (see \Cref{s:mf_P}). They can be computed given either an input angular power spectrum or $\mu$. 
    \item To determine the MFs on any scalar map, either on the full sky or within any mask.
\end{enumerate}

Both $T$ and $P^2$ are scalar fields on the sphere, and therefore the MFs can be computed with the same code for both quantities, although their expected results are different. We describe the implementation in this section, where we refer generally to an arbitrary (smooth) field on the sphere $f(x)$. We compute the MFs normalised by the area of the sphere ($v_i = \frac{V_i}{4\pi}$), so the extension to masked maps is straight--forward.

Without loss of generality, for this work we normalise input maps to have unit variance, as we did in the previous sections when computing the theoretical predictions. In the case of $P^2$, both the $U$ and $Q$ maps are separately normalised to unit variance. The MFs can now be expressed as a function of the a-dimensional threshold $u$.

\subsection{Computation of the MFs}

\subsubsection*{First MF, $\mathbf{v_0}$} 

The first MF, $v_0$, is computed according to the equation:
	\begin{equation}
		v_0(u) = \frac{1}{4\pi} \int_{\mathbb{S}^{2}} \Theta(f(x)-u) \, dx
	\end{equation}
	where $\Theta(r)$ is the Heaviside function ($1$ where $r\geq0$; $0$ otherwise). In the computation, we approximate the integral with a sum over all valid pixels. The integral divided by the area of the sphere, $4\pi$, can be seen as the average value of the argument inside the integral. Therefore, it can be approximated with the average over all pixels. Thus, $v_0(u)$ is computed as the fraction of the map above the threshold $u$.
 
\subsubsection*{Second MF, $\mathbf{v_1}$} 
The second MF, $v_1$, is computed according to the equation:
	\begin{align}
		\begin{split}
			v_1(u) &=\frac{1}{4\pi} \frac{1}{4} \int_{\partial A_{u}}dr \\
			&=  \frac{1}{4\pi} \frac{1}{4}  \int_{\mathbb{S}^{2}} \delta(f(x)-u) \cdot |\nabla f| \, dx
		\end{split}
	\end{align}
	where the second equality is the result of a change of coordinates from the line element $dr$ of the boundary of the excursion set to a surface element $dx$ on the sphere. The gradient of $f$ is denoted by $\nabla f$ and $\delta$ is the Dirac delta.
	
	We start by computing $|\nabla f|$ at every pixel. Then, in order to approximate the integral with a sum over pixels, we provide a bin of $\Delta u$. We create a mask $M_u(x)$ identifying the pixels where the value of the map is close to $u$:
	\begin{equation}
		M_u(x) =
		\begin{cases}
			1  & \text{if } u-\frac{\Delta u}{2} \leq f(x) \leq u+\frac{\Delta u}{2} \\
			0 & \text{otherwise}
		\end{cases}
	\end{equation}
	Now, $v_1(u)$ can be computed as the mean of the pixels of ${M_u \cdot |\nabla f|}$, divided over $\Delta u$, the bin size, and $4$, the normalisation factor.

\subsubsection*{Third MF, $\mathbf{v_2}$} 
The third MF, $v_2$, is computed according to the equation:
	\begin{align}
		\begin{split}
			v_2(u) &= \frac{1}{4\pi}\frac{1}{2\pi }\int_{\partial A_{u}}\kappa (r)dr \\
			&=  \frac{1}{4\pi}\frac{1}{2\pi}\int_{\mathbb{S}^{2}} \delta(f(x)-u) \cdot |\nabla f| \cdot \kappa \, dx
		\end{split}
	\end{align}
	The change of coordinates to surface element and the approximation to a finite sum of pixels is analogous to the previous case. We define the mask $M_u$ identically and compute the mean of the pixels of $M_u \cdot |\nabla f| \cdot \kappa$. This quantity is then divided by the bin size $\Delta u$ and by $2\pi$ to yield the result for $v_2(u)$. 
	
	In this formula, $\kappa$ is the geodesic curvature of the boundary of the excursion set. Given that this curve is given by the implicit equation $f(x)=u$, we can use the expression of $\kappa$ for an implicit curve \citep[see, e.g.,][]{docarmo1976}:
	\begin{equation}
		\kappa = \frac{2f_{;\phi} f_{;\theta} f_{;\phi \theta}-f_{;\theta}^2 f_{;\phi\phi}-f_{;\phi}^2 f_{;\theta\theta}}{\left(f_{;\phi}^2+f_{;\theta}^2\right)^\frac32}
	\end{equation}
	where the semicolon denotes the covariant derivative. All the quantities involved in this expression can be computed for each pixel.

\subsection{Computation of the derivatives}
The spatial derivatives of the maps (both partial and covariant) are computed with the help of the healpy\footnote{\url{https://github.com/healpy/healpy}} function \texttt{alm2map\_der} in harmonic space \citep{Zonca2019}. 
The covariant derivatives of a field $f$ on the sphere are computed in terms of the partial derivatives:
\begin{align*}
f_{;\theta} &=f_{,\theta}\\
f_{;\phi} &= \frac{f_{,\phi}}{\cos{\theta}} \\
f_{;\theta\theta} &=f_{,\theta\theta}\\
f_{;\theta\phi} &= \frac{f_{,\theta\phi}}{\cos{\theta}}-\frac{\sin{\theta}\cdot f_{,\phi}}{\cos^{2}{\theta}}
\\
f_{;\phi\phi} &= -\frac{f_{,\phi\phi}}{\cos^{2}{\theta}} + f_{,\theta}\cdot\tan{\theta}
\end{align*}%
where we consider $\theta\in(-\pi/2, \pi/2)$ as the latitude and $\phi\in[0,2\pi)$ as the longitude.

The need of estimating the spatial derivatives of the map means that the computations are reliable only for maps whose angular power spectrum is negligible at multipoles close to $\ell_{max}=3\cdot N_{\textrm{side}} -1$; i.e., the derivatives can only be computed if the maps are reasonably smooth. This becomes a requirement in order to estimate $v_1$ and $v_2$, while $v_0$ can be computed for any map as it does not require the computation of the derivatives.

This smoothness condition is satisfied for realistic $T$ and $P^2$ maps, as they are reasonably smooth, but it can break if one tries to analyse maps of pure noise or heavily dominated by the modes at the highest multipoles (corresponding to pixel--size scales). We note that our theoretical results only hold for functions that are at least twice--differentiable almost everywhere \citep[for other cases, such as a fractal behaviour, see][]{lan2018}.

\section{Simulations and Data}
\label{s:data}

\subsection{Gaussian simulations}
\label{ss:simsgau}
In order to validate the implementation of the computation on HEALPix maps and the theoretical predictions, we generate $300$ Gaussian isotropic CMB maps of the Stokes parameters (Q, U). We use the healpy function \texttt{synfast} with the best fit angular power spectrum for polarization (E and B modes) as reported by \citet{planck_spectra}. For these simulations, we fix a resolution of $N_{\textrm{side}}=1024$; this is half of the \textit{Planck} value in order to reduce the computational time without a significant loss of resolution; we have verified that the results remain unchanged by this choice.

For the estimation of $v_{0}$, we exclude the modes with $\ell \leq 4$ in the simulated maps in order to avoid spin effects in the computation of the MF (see \Cref{s:mf_P}). Conversely, when $v_1$ and $v_2$ are estimated, a smoothing of the maps with a Gaussian beam with full width at half maximum (FWHM) equal to $15^{\prime}$ is applied in order not to be affected by pixelisation effects in the computation of the spatial derivatives.

MFs have been computed over the full sky maps of $P^2 = Q^2 + U^2$, where $Q$ and $U$ have been previously normalised to unit variance. We analyse the deviation of each MF at different values of the threshold $u$ in these simulated realisations of Gaussian isotropic CMB with respect to the theoretical expectations (see \Cref{ss:validation}). 
In this way, we are able to simultaneously assess the accuracy of the formulae obtained in \Cref{s:mf_P} and verify the implementation of MFs computations on simulated maps. 

\subsection{\textit{Planck} polarization maps}
\label{ss:planckmaps}
After validating the machinery with Gaussian isotropic CMB simulations, we apply the MFs to \textit{Planck} CMB polarization data. In order to verify that the results are consistent and do not depend on the component separation algorithm used to reconstruct the CMB, we use the polarization maps released by the \textit{Planck} collaboration obtained with two different pipelines: SMICA (Spectral Matching Independent Component Analysis) and SEVEM (Spectral Estimation Via Expectation Maximization), as reported in \citet{2020A&A...641A...4P}.

SMICA \citep{smica} performs a foregrounds and noise subtraction by linearly combining \textit{Planck} frequency maps in harmonic space with multipole-dependent weights, up to $\ell=4000$, which minimize on each angular scale the variance of the output map. In polarization, it combines the $E$ and $B$ modes independently and then recombines them to obtain the $Q$ and $U$ maps. 

SEVEM \citep{smica1}, instead, is a template-fitting method. Different templates of the foregrounds emission are estimated as the difference of maps from neighbouring frequencies. This procedure removes the CMB signal while preserving the Galactic one. A linear combination of these templates is then subtracted from the input data to produce a cleaned CMB map at a specific \textit{Planck} frequency. The coefficients of the linear combination are determined by minimising the variance of the cleaned map outside a mask, which covers the point sources detected in polarization and the $3\%$ brightest Galactic emission. The $100$, $143$, and $217$ GHz cleaned maps are then combined in harmonic space, using E and B decomposition, to produce the final CMB maps for the Q and U components at a resolution of FWHM=$5'$ and a maximum considered multipole of $\ell=3000$.

The native resolution of these released maps\footnote{\url{pla.esac.esa.int/pla/}\label{ft:pla}} is $N_{\textrm{side}}=2048$; we degrade them to $N_{\textrm{side}}=1024$ to match exactly the pipeline applied for simulations. Similarly, before the computation of $V_0$ we again exclude the lowest multipoles ($\ell \leq 4$) in order to avoid spin effects, while for the computation of $V_1$ and $V_2$ we bring the maps to a resolution of FWHM=$15^{\prime}$ to not be affected by pixelization effects. 

\begin{figure}
	\centering
	\includegraphics[width=0.45\textwidth]{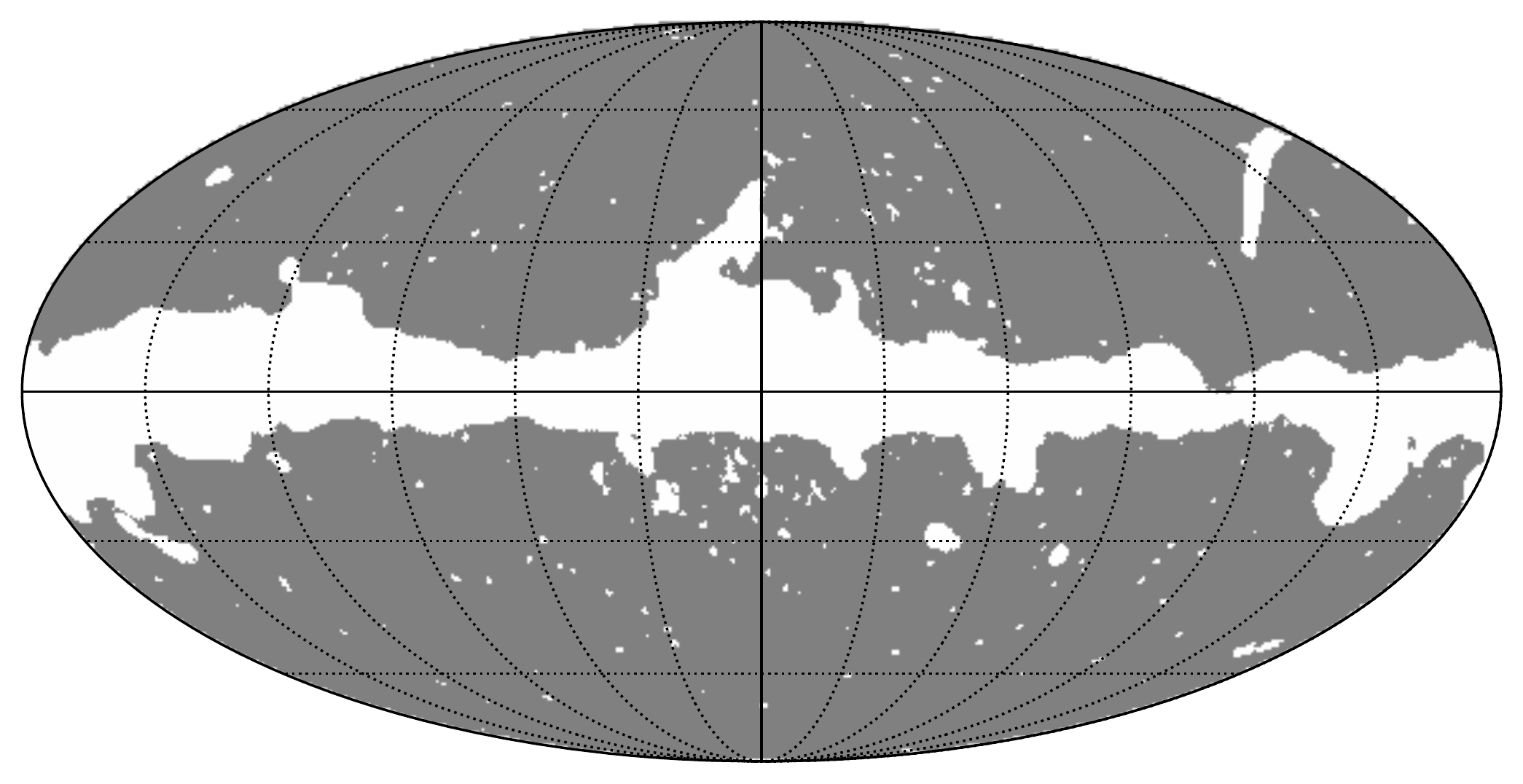} 
\caption{2018 Component Separation common \textit{Planck} mask for polarization. It excludes the most contaminated regions near the Galactic plane and the polarized point sources. It keeps a total sky fraction of $f_{sky}=78\%$.}
\label{fig:Planck mask}
\end{figure}

\begin{figure}
	\centering
	\includegraphics[width=0.45\textwidth]{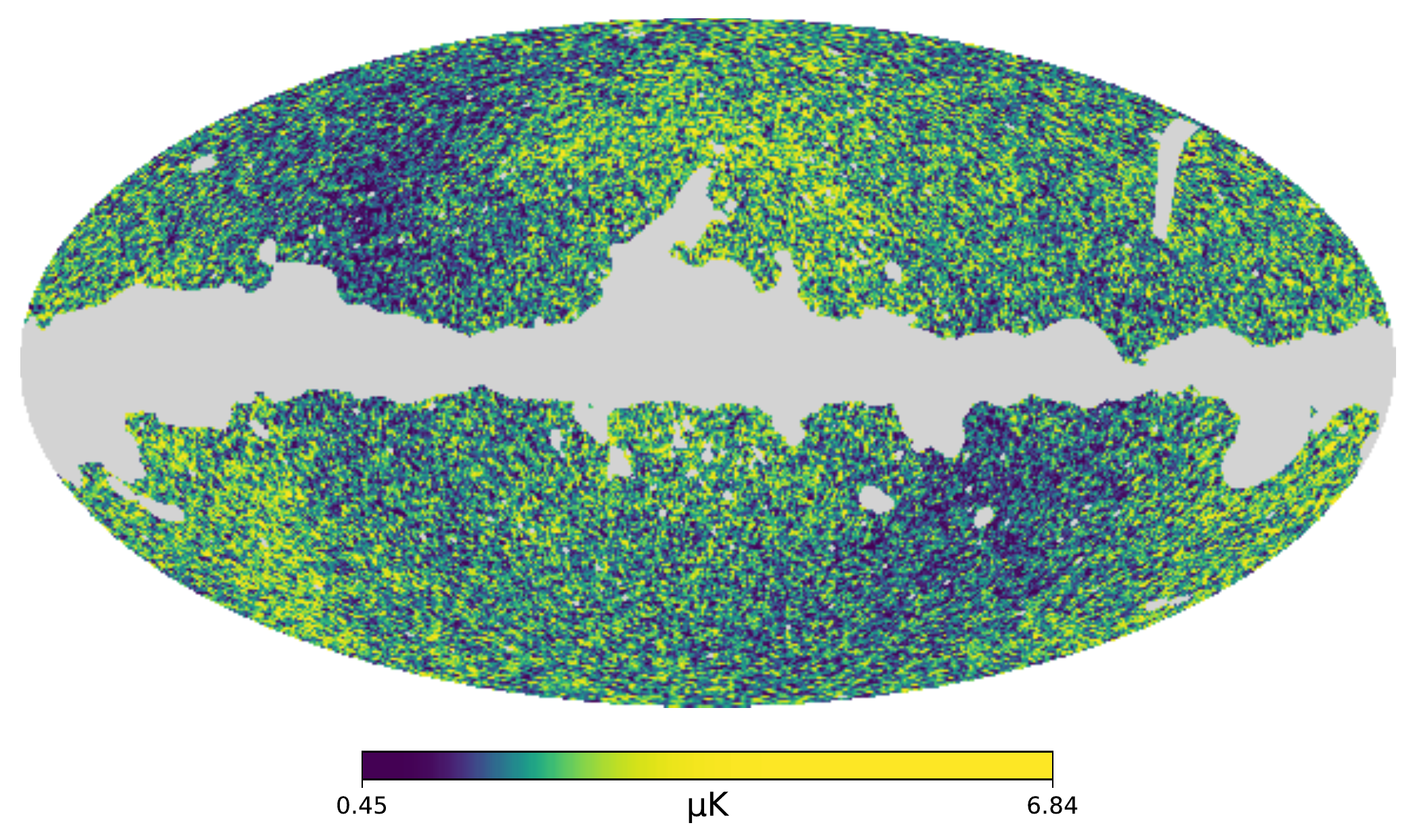} 
	\includegraphics[width=0.45\textwidth]{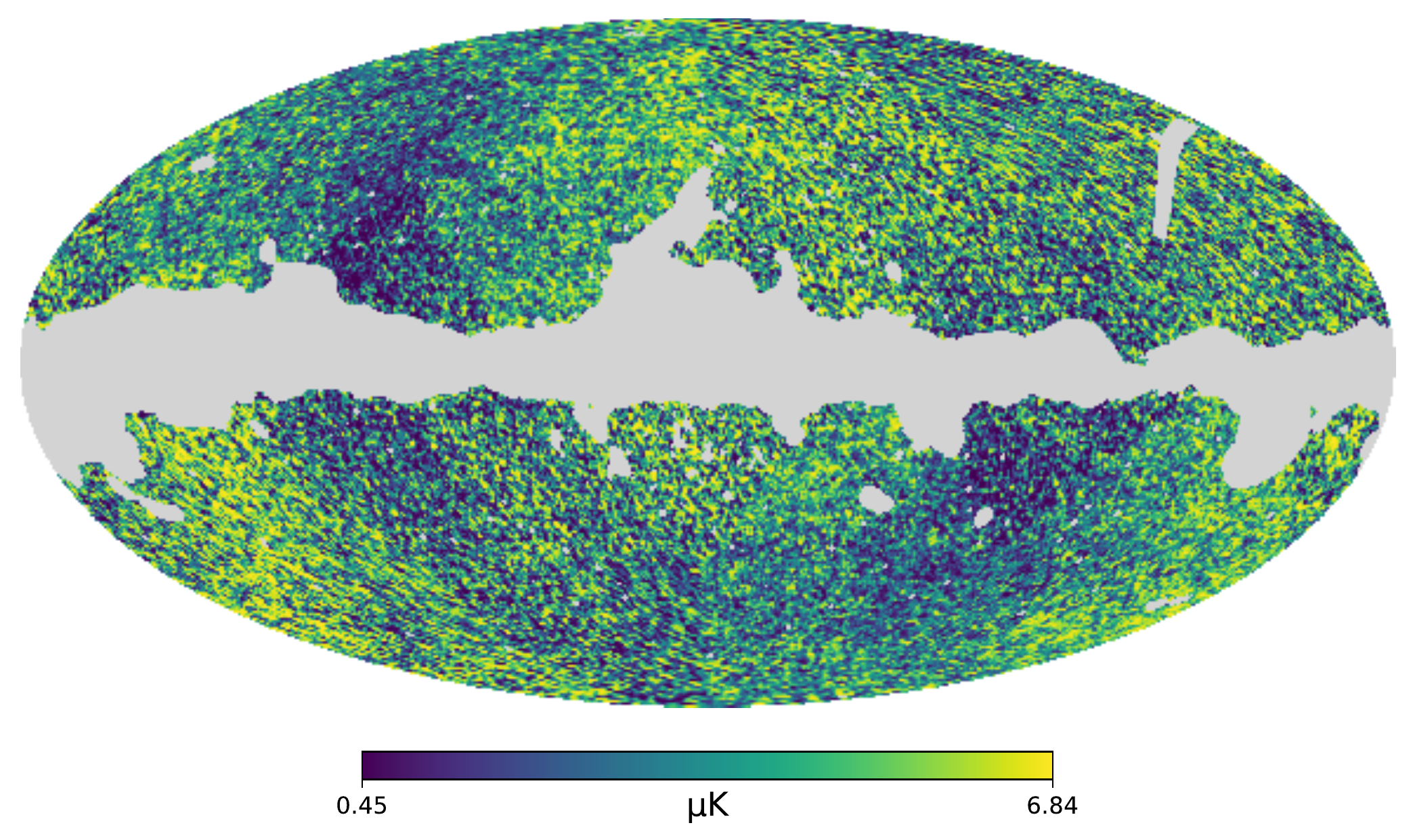}
\caption{\textit{Planck} CMB maps for $P=\sqrt{Q^2 + U^2}$ obtained with the SMICA (top) and SEVEM (bottom) pipelines. They have been smoothed with a Gaussian beam of FWHM=$30'$ for visualisation purposes.}
\label{fig:Planck maps}
\end{figure}

\textit{Planck} CMB polarization maps still present residual contamination due to Galactic foregrounds and point sources. The most contaminated regions of the sky must be excluded from any cosmological analysis. Therefore, we adopt the 2018 Component Separation Common polarization mask (shown in \Cref{fig:Planck mask}) throughout all the analysis, as recommended by \citep{2020A&A...641A...4P}. This mask was obtained by thresholding the standard deviation between each of the four cleaned \textit{Planck} CMB maps and it is then augmented with the Commander and SEVEM confidence masks, as well as with the SEVEM and SMICA in-painting masks. 
The masked SMICA and SEVEM $P$ maps are shown in \Cref{fig:Planck maps}. 

\subsection{\textit{Planck} end--to--end simulations}
\label{ss:plancksims}
The SMICA and SEVEM maps include residual effects associated to \textit{Planck} instrumental noise, which are especially relevant in polarization. Due to the scanning strategy of the satellite, this component presents an anisotropic pattern. Such contamination has to be properly taken into account for any cosmological analysis of the \textit{Planck} CMB maps. To this end, the \textit{Planck} Collaboration released end--to--end simulations\textsuperscript{\ref{ft:pla}} which include the expected CMB signal and noise residuals. 

These simulations have been obtained by generating realistic CMB (Gaussian and isotropic) and noise maps at the different \textit{Planck} frequency channels, which are then processed by each component separation algorithm. The resulting maps are thus able to reflect the expected statistical properties across the sky of noise residuals and CMB given by each method. 

We compare the MFs of the SMICA and SEVEM released CMB polarization maps with those computed on these realistic simulations. Compatible results would suggest that any non--Gaussianity or deviation from statistical isotropy present in CMB polarization maps is compatible with the ones produced by noise residuals. A significant deviation might signal the presence of residual foregrounds, unmodelled systematic effects, or a hint of non--Gaussianity or departure from statistical isotropy of the CMB with primordial origin.

We use both SMICA and SEVEM simulations with $N_{\textrm{side}}=1024$ to make all the analysis identical in all maps. Similarly to real \textit{Planck} polarization maps, we again exclude the lowest multipoles ($\ell \leq 4$) before the computation of $V_0$, while we bring the maps to a resolution of FWHM=$15^{\prime}$ for the computation of $V_1$ and $V_2$.

\subsection{Needlet maps}
\label{ss:needlets}

The comparison of MFs computed on \textit{Planck} $P^2$ maps with those of end-to-end simulations allows us to perform a statistical analysis over the full range of angular scales observed by the \textit{Planck} satellite. 
However, any non-Gaussian or statistically anisotropic signal may be relevant only in a specific range of multipoles (e.g. Galactic residuals on large scales, point sources at high multipoles). Therefore, assessing the compatibility between maps and end-to-end simulations separately at different angular scales may provide a more detailed statistical inspection of the data. We thus compare MFs of \textit{Planck} SMICA and SEVEM $P^2$ maps with those of realistic end-to-end simulations (introduced, respectively, in \Cref{ss:planckmaps} and \Cref{ss:plancksims}) in needlet domain. A similar analysis for $E$-mode \textit{Planck} maps has been reported in \citet{2020A&A...641A...7P}.

Needlets are a wavelet system which guarantees simultaneous localisation in harmonic and pixel space; they have been introduced in the statistical literature by \citet{narcowich2006} and firstly applied to CMB data by \citet{2006PhRvD..74d3524P}. Given any field of spin $s$ defined on the sphere, the corresponding needlet maps $\beta^{s}_{j}$ are obtained by filtering its harmonic coefficients $a^{s}_{\ell m}$ with a harmonic weighting function $b_{j}(\ell)$ which isolates modes at different angular scales for each needlet scale $j$ \citep{geller2010spin}:
\begin{equation}
\beta^{s}_{j}(x)
=\sum_{\substack{\ell,m}} (a^{s}_{\ell m}\cdot b_{j}(\ell))\cdotp _{s}Y_{\ell m}(x),
\label{eq:needlets_map}
\end{equation}
where $x$ is a direction in the sky and $_{s}Y_{\ell m}$ are the spin-weighted spherical harmonics. Such procedure in harmonic space is equivalent to performing a convolution of the map in real domain. The shape of the needlet bands is defined by the choice of the harmonic function $b$ whose width is set by the parameter $B$: lower values of $B$ correspond to a tighter localisation in harmonic space (less multipoles entering into any needlet coefficient), whereas larger values ensure wider harmonic bands. 
\begin{figure}
	\centering
    \includegraphics[width=0.48\textwidth]{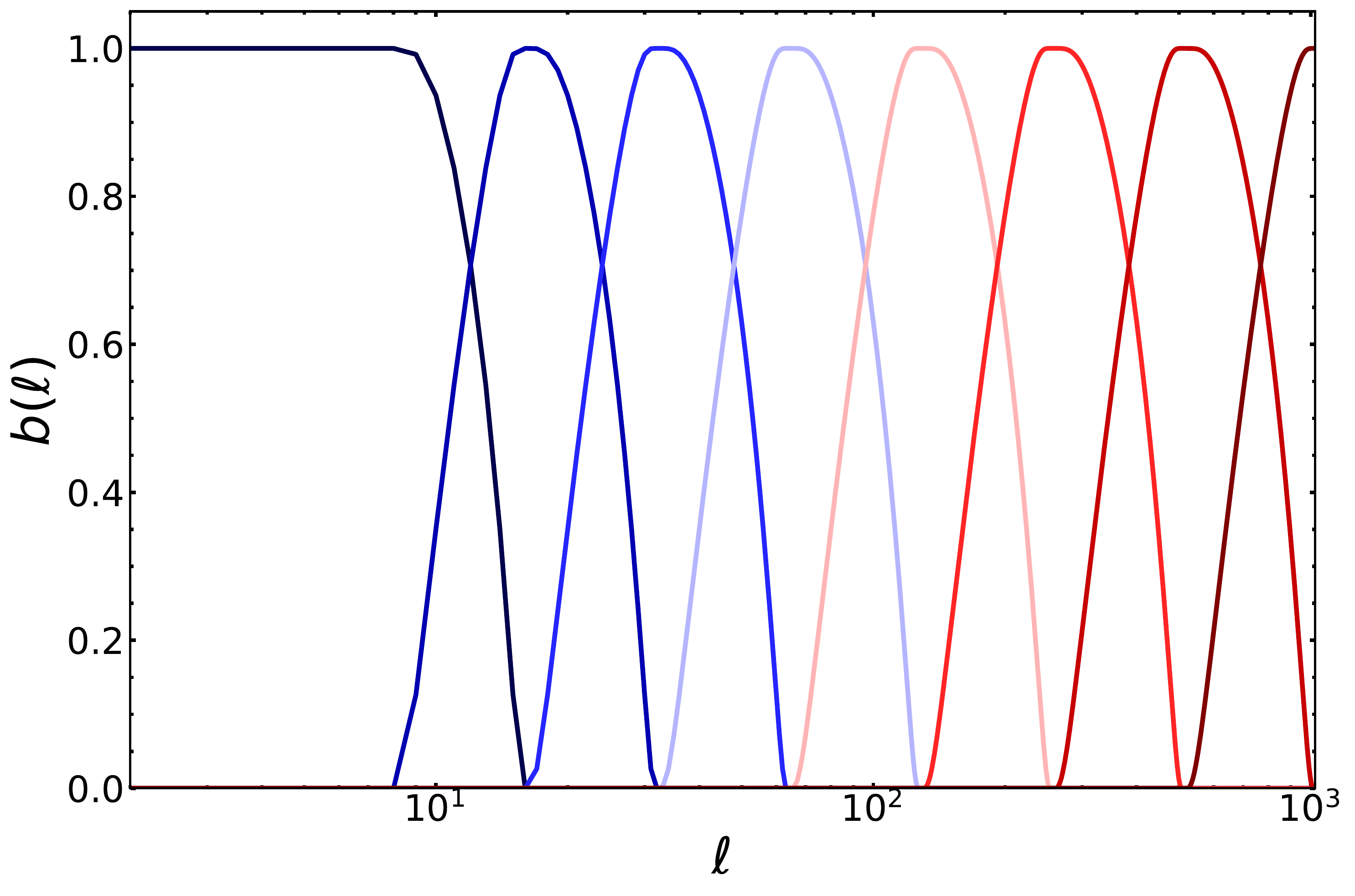} 
	\caption{Representation in the harmonic domain of the needlet configuration adopted in this analysis: standard needlets with $B=2$. The first four bands have been merged together according to \cref{eq:merge}, while the other bands are left unchanged. }
	\label{f:bands_nl}
\end{figure}

As already mentioned in \Cref{s:mf_P}, the Stokes parameters $Q$ and $U$ are components of a spin-2 field on the sphere:
\begin{equation}
Q \pm i U(x)
=\sum_{\substack{\ell,m}} a^{\pm 2}_{\ell m}\cdotp _{\pm 2}Y_{\ell m}(x).
\label{eq:QU_spin2}
\end{equation}
Therefore, in this paper, we first filter the $a^{\pm 2}_{\ell m}$ coefficients to obtain $Q$ and $U$ needlet maps and then compute the corresponding $P^2$ map. As needlet harmonic function $b$, we have adopted the standard needlet construction \citep{narcowich2006} with $B=2$. The needlet filters have been obtained through the Python package \textit{MTNeedlets}\footnote{\url{https://javicarron.github.io/mtneedlet}} \citep{carronduque2019}. 

Since the first needlet bands would select only very few large-scale modes, we merge together the first four bands as follows:
\begin{equation}
b_{0}^{\textrm{new}}(\ell) =\sqrt{\sum_{j=0}^{3}b_{j}^{2}(\ell)}.
\label{eq:merge}
\end{equation}
The adopted configuration of needlet harmonic filters is shown in \Cref{f:bands_nl}. We can note that the $j$-th needlet band is non-zero only for $\ell \le 2^{j+4}$. Therefore needlet maps are band-limited and it suffices to reconstruct them at $N_{\textrm{side}} = 2^{j+3}$. As it was done in \citet{2020A&A...641A...7P}, at each needlet scale, corresponding masks are obtained by first downgrading the mask at the targeted resolution as an intensity map. The resulting smoothed downgraded mask is then thresholded by setting pixels with values lower than $0.9$ to zero and all others to unity, so that we again have
a binary mask. 

\subsection{\textit{LiteBIRD} and SO simulations}
\label{ss:LB_SO_sims}
CMB intensity anisotropies have been measured by \textit{Planck} with very high accuracy \citep{planck_spectra}. Conversely, polarization data are mostly noise-dominated and contaminated by several instrumental systematic effects. As already discussed in \Cref{s:intro}, several experiments, either from the ground or from space, are planned to measure CMB polarization with unprecedented sensitivity. Therefore, in this paper, we forecast the capability of MFs of the polarization modulus to detect a deviation from Gaussianity and statistical isotropy in realistic simulated maps for two future and complementary experiments: SO \citep{2019JCAP...02..056A} and the \textit{LiteBIRD} satellite \citep{PTEPlitebird}. The SO is composed of a Large Aperture Telescope (SO-LAT) and several refracting Small
Aperture Telescopes (SO-SATs). SO-LAT will observe $40\%$ of the sky and will be devoted to measurements of the CMB small-scale
temperature and polarization anisotropies, the CMB lensing, the SZ effects, and extragalactic sources. SO-SATs, instead, aim at a detection of primordial $B$ modes. Given the affinity with the scientific target of \textit{LiteBIRD}, in this analysis we have considered only SO-SATs. 

For both experiments, we simulate a fiducial Gaussian data set and a test data set which also includes a non-Gaussian anisotropic signal. In the fiducial Gaussian data set, $Q$ and $U$ maps are obtained by adding Gaussian and isotropic CMB and noise realisations:
\begin{equation}
    \{Q,U\}_{\textrm{fid}} = \{Q,U\}_{\textrm{CMB}} + \{Q,U\}_{\textrm{noise}}. 
\end{equation}
In the test data set, instead, we also add a map of simulated Galactic foreground emission as an example of a generic non-Gaussian and anisotropic component:
\begin{equation}
    \{Q,U\}_{\textrm{test}} = \{Q,U\}_{\textrm{CMB}} + \{Q,U\}_{\textrm{noise}} + \alpha_{\textrm{f}}\cdot \{Q,U\}_{\textrm{fgd}}, 
\label{e:test_maps}
\end{equation}
where $\alpha_{\textrm{f}}$ is a scaling factor which modulates the overall amplitude of this non-Gaussian contamination. Such rescaling roughly parametrizes the level of residual Galactic foregrounds after a component separation analysis has been performed on CMB multi-frequency data. We let $\alpha_{\textrm{f}}$ to vary in order to test the sensitivity of $P^2$ MFs to different magnitudes of the simulated non-Gaussian and anisotropic signal across the sky. 

The CMB maps are generated with the HEALPix Python package \citep{Zonca2019} from the angular power spectra of the best-fit \textit{Planck} 2018 parameters \citep{2020A&A...641A...6P}. For the instrumental noise component, we first simulate noise maps at the different frequency channels of the considered experiment as Gaussian realisations from the following power spectrum:
\begin{equation}
    N_{\ell} = \left(\sigma_{\textrm{Q,U}} \cdot \frac{\pi}{180\cdot 60}\right)^2 \cdot  \left[1+\left(\frac{\ell}{\ell_{\textrm{knee}}}\right)^{\alpha_{\textrm{knee}}}\right],
\end{equation}
where $\sigma_{\textrm{Q,U}}$ is the polarization sensitivity in $\mu \textrm{K}_{\textrm{CMB}}\cdot$arcmin, while $\ell_{\textrm{knee}}$ and
$\alpha_{\textrm{knee}}$ refer to the contribution from 1/f noise. 
The employed \textit{LiteBIRD} and SO-SATs specifications are taken from the latest forecasts of the two experiments \citep{PTEPlitebird,2019JCAP...02..056A}. 
For \textit{LiteBIRD}, we do not consider the 1/f term whose contribution is expected to be fully suppressed by the modulation of the half wave plate. For SO-SATs we have adopted the values for $\ell_{\textrm{knee}}$ and
$\alpha_{\textrm{knee}}$ reported in \citet{2019JCAP...02..056A} for the pessimistic 1/f case. In this way, we will obtain the most conservative constraints on the detection power of $P^2$ MFs for SO-SATs, which is likely to improve for more realistic instrumental configurations.
\begin{figure}
\centering      
\includegraphics[width=0.48\textwidth]{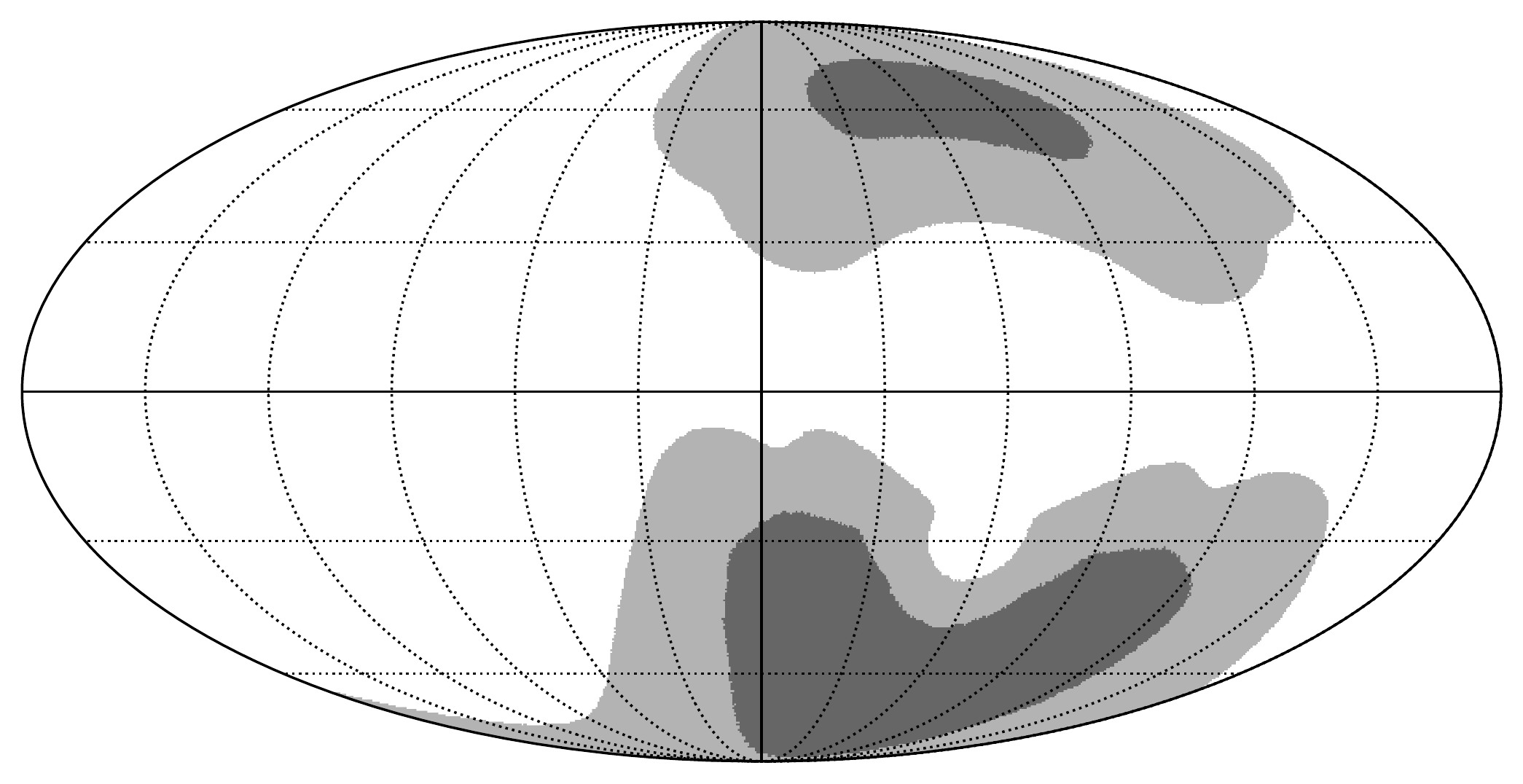}
    \caption{In light grey: The SO-SATs full footprint with sky fraction $f_{sky}=34\%$. In dark grey: The SO-SATs patch considered in this analysis, which corresponds to the sky area with the highest signal-to-noise ratio in the full footprint. This was obtained considering pixels with the largest values in the map of hit counts for a final sky fraction $f_{sky}=10\%$.}
\label{fig:patch_so}
\end{figure}
For both experiments, we neglect for simplicity the scanning strategy of the telescopes and, thus, the statistical properties of the noise are assumed constant across the sky. This is a fair assumption for \textit{LiteBIRD}, since, as also discussed in \citet{PTEPlitebird}, the scanning strategy parameters are chosen so as to have the distribution of hit counts as uniform as possible and are expected to have a completely negligible effect in $P^2$ maps. For SO-SATs, this statement does not hold, because the hit counts map is very inhomogeneous. Therefore, we restrict the MFs computation to a smaller patch with $f_{\textrm{sky}}=10\%$ (shown in \Cref{fig:patch_so}), where the scanning strategy can be confidently assumed uniform. Such patch is obtained by considering only the $10\%$ of the pixels with the highest values in the hit counts map provided by the SO Collaboration. 
The adopted SO-SATs sensitivities, which are reported in \citet{2019JCAP...02..056A}, indeed refer to a homogeneous hit counts map with such a sky fraction. 
The final noise map, for both experiments, is obtained by linearly combining the central frequency channels noise maps with an inverse noise weighting. We considered the frequency range $93 \le \nu \le 225$ for SO-SATs and $89 \le \nu \le 235$ for \textit{LiteBIRD}.

In this paper, we aim at providing a first simplified forecast of the capability of the MFs to detect non-Gaussian and anisotropic signals in $P^2$ maps for experiments with much higher sensitivity with respect to \textit{Planck}. A more detailed analysis would require considering noise residuals given by component separation pipelines. However, we do not want to add the additional degree of freedom of the choice of specific component separation methods to be applied to the considered experiments and we leave such analysis for future work.

The non-Gaussian and anisotropic signal $\{Q,U\}_{\textrm{fgd}}$ injected in the test data set of \cref{e:test_maps} is a simulated polarized Galactic foreground emission at $100$ GHz multiplied by a varying rescaling factor $\alpha_{\textrm{f}}$. This signal is generated with the \textit{PySM} Python package\footnote{\url{https://github.com/galsci/pysm}} \citep{2017MNRAS.469.2821T} and includes the two main polarized Galactic components: synchrotron and thermal dust emission. They are, respectively, modelled with the \textit{PySM} \texttt{s5} and \texttt{d10}. The choice of these models is driven by the fact that non-Gaussianity is injected in them up to much smaller scales than in other analogous commonly used \textit{PySM} models as \texttt{s1} and \texttt{d1}. The synchrotron 
Spectral Energy Distribution (SED) follows a power law:
\begin{equation}
	\{Q,U\}_{\textrm{s}}(x,\nu) = \{Q,U\}_{\textrm{s}}(x,\nu_{0}) \cdot  \Bigg(\frac{\nu}{\nu_{0}} \Bigg)^{\beta_{\textrm{s}}(x)}, 
\label{eq:sync}
\end{equation}
where $\beta_{\textrm{s}}$ is the spectral index, $x$ the position in the sky, and $\nu$ the considered frequency. $\{Q,U\}_{\textrm{s}}(x,\nu_{0})$ represents the synchrotron template at a reference frequency $\nu_{0}$ and is the \textit{WMAP} $9$-year $23$-GHz $Q$/$U$ map \citep{2013ApJS..208...20B} smoothed with a Gaussian kernel of FWHM$=5^\circ$ with non-Gaussian small scales added through the Logarithm of the Polarization Fraction Tensor (\texttt{logpoltens}) formalism. 
The thermal dust 
SED is assumed to be a modified black-body (MBB):
\begin{equation}
	\{Q,U\}_{\textrm{d}}(x,\nu) = \{Q,U\}_{\textrm{d}}(x,\nu_{0}) \cdot  \Bigg(\frac{\nu}{\nu_{0}} \Bigg)^{\beta_{\textrm{d}}(x)} \cdot \frac{B_{\nu}(T_{\textrm{d}}(x))}{B_{\nu_{0}}(T_{\textrm{d}}(x))},
	\label{eq:dust}
\end{equation}
where $B_{\nu}(T)$ is the black-body spectrum, $\beta_{\textrm{d}}$ is the dust spectral index, $T_{\textrm{d}}$ is the dust temperature, and $\{Q,U\}_{d}(x,\nu_{0})$ represents the dust template at a reference frequency $\nu_{0}$. Both the spectral parameters maps and the template are obtained from the application of the Generalised Needlet Internal Linear Combination (GNILC) to \textit{Planck} data \citep{2020A&A...641A...4P} and non-Gaussian small-scale fluctuations have been added in the \texttt{logpoltens} formalism. 

We note that the injected non-Gaussian anisotropic signal is particularly relevant on large and intermediate angular scales. We thus expect to be able to constrain effectively the detection power of the proposed tools only in that range of multipoles, which, however, is exactly that targeted by \textit{LiteBIRD} and SO-SATs for accurate measurements of the polarization anisotropies. Moreover, small scales are usually noise-dominated. Therefore Galactic emission represents a good probe to forecast the MFs detection power for these future CMB experiments. If one wants to expand the assessment of MFs sensitivity at smaller scales, they would have to choose a signal model different than Galactic emission with non-Gaussianities and anisotropies at those scales.

The capability of MFs to detect deviations from Gaussianity and statistical isotropy in the test data set for SO-SATs and \textit{LiteBIRD} is compared with that of \textit{Planck}. We thus generated also for \textit{Planck} a fiducial and a test data set, where CMB and Galactic foregrounds maps are the same of \textit{LiteBIRD} and SO-SATs, while, as noise component, we used the end-to-end SMICA noise residuals maps introduced in \Cref{ss:plancksims}. All maps are brought to the resolution of $37.8^{\prime}$ for the comparison between \textit{Planck} and \textit{LiteBIRD}, of $30^{\prime}$ for the comparison between \textit{Planck} and SO-SATs.

\subsection{Compatibility analysis}
\label{ss:chi}
In order to assess the level of deviation between MFs on different maps, we use the $\chi^2$ statistic. Let $X$ be a vector with the result of any of the MFs with the mean of the simulations subtracted and $\Sigma$ the covariance matrix of $X$ (representing the correlation of MFs at different thresholds). The $\chi^2$ statistic is given by:
\begin{equation}
	\chi^2 = X^t \Sigma^{-1} X
\label{e:chi2}
\end{equation}

If $S$ is the sample covariance (computed with simulations), an unbiased estimator of the inverse of the covariance is:
\begin{equation}
	\Sigma^{-1} = \frac{N-d-2}{N-1} S^{-1}
\label{e:inv_covariance}
\end{equation}
where $N=300$ is the number of simulations and $d=28$ is the number of bins in the considered range of thresholds $u$. The pre-factor in this equation is known as the Hartlap factor \citep{hartlap2007} and it is needed for the estimator to be unbiased. 

In this analysis we report the reduced $\hat{\chi}^2=\chi^2/N_{\textrm{dof}}$, where $N_{\textrm{dof}}=28$ is the number of degrees of freedom (\textit{i.e}, the number of thresholds in our computation). We also compute the probability--to--exceed a given $\chi^2$:
\begin{equation}
    p_{\textrm{exc}}(\chi^2_{\textrm{data}}) = P(\chi^2_{\textrm{sims}} > \chi^2_{\textrm{data}})
    \label{eq:chi2_P2}
\end{equation}
where $\chi^2_{\textrm{data}}$ and $\chi^2_{\textrm{sims}}$ are respectively associated to any MF computed on the data or on a simulation. In \cref{eq:chi2_P2}, a value close to $50\%$ means that the data closely resemble the expectation, while a value close to $0\%$ or $100\%$ means that the data are highly incompatible with simulations.

Finally, we also report the significance of the deviation $\Delta_{\chi}$ of $\chi^2$ in the data with respect to the mean $\bar{\chi}^2_{\textrm{sims}}$ among the simulations, reported in units of standard deviation $\sigma_{\textrm{sims}}$ (see last column of Table \ref{t:chiscmb}):
\begin{equation}
	\Delta_{\chi} = \frac{\chi^2_{\textrm{data}}-\bar{\chi}^2_{\textrm{sims}}}{\sigma_{\textrm{sims}}}.
\label{eq:delta_chi}
\end{equation}
To assess the $P^2$ MFs detection power of a non-Gaussian anisotropic signal in the simulated data sets introduced in \Cref{ss:LB_SO_sims}, we compute a $\chi^2$ for each realisation as in \cref{e:chi2}, where $X$ is the MFs difference between the test and the fiducial Gaussian data set for that realisation and the covariance matrix is estimated from MFs of the fiducial data set. We then report for each case the average probability for a $\chi^2$ distribution with $N_{\textrm{dof}}=28$ (the adopted number of thresholds) to exceed the computed $\chi^2$s. If this value is close to $50\%$, it means that no deviations are observed between MFs of the test and Gaussian fiducial data set. Conversely, if such probability is very close to $0\%$, we are able to detect at high significance the injected non-Gaussian anisotropic signal in the test data set.

\section{Results}
\label{s:results}
In this section, we proceed to validate the code and the theoretical formulae for the expected values of the MFs for $P^2$ maps. In order to assess the capability of these tools to detect deviations from Gaussianity and statistical isotropy, we then apply the pipeline to \textit{Planck} CMB polarization data and simulated $P^2$ maps of \textit{LiteBIRD} and SO. We use $300$ simulations for each case. 

Throughout all the following analysis, we set thresholds between $u=0$ and $5$. For visualisation purposes, all MFs are plotted in $100$ bins, which correspond to a spacing of $\Delta u=0.05$. The statistical analysis described in \Cref{ss:chi} is instead performed lowering the number of bins to $28$ ($\Delta u=0.18$) so to get a more robust estimate of the covariance matrix given in \cref{e:inv_covariance}.

In \Cref{ss:validation}, we compare the theoretical expectation of MFs for a $P^{2}=Q^{2}+U^{2}$ field under the assumption of Gaussianity and isotropy with those computed on CMB simulations generated with the same statistical properties. In \Cref{ss:planck_analysis}, MFs are estimated on the \textit{Planck} SMICA and SEVEM CMB polarization modulus map and compared with those of the corresponding end-to-end simulations. In \Cref{ss:fcasts}, we assess the detection power of a non-Gaussian and anisotropic signal at different angular scales for MFs computed on realistic simulated $P^2$ data sets of \textit{LiteBIRD} and SO and the results are compared with those obtained from analogous \textit{Planck} simulations.
\begin{figure}
	\centering
	\includegraphics[width=0.47\textwidth]{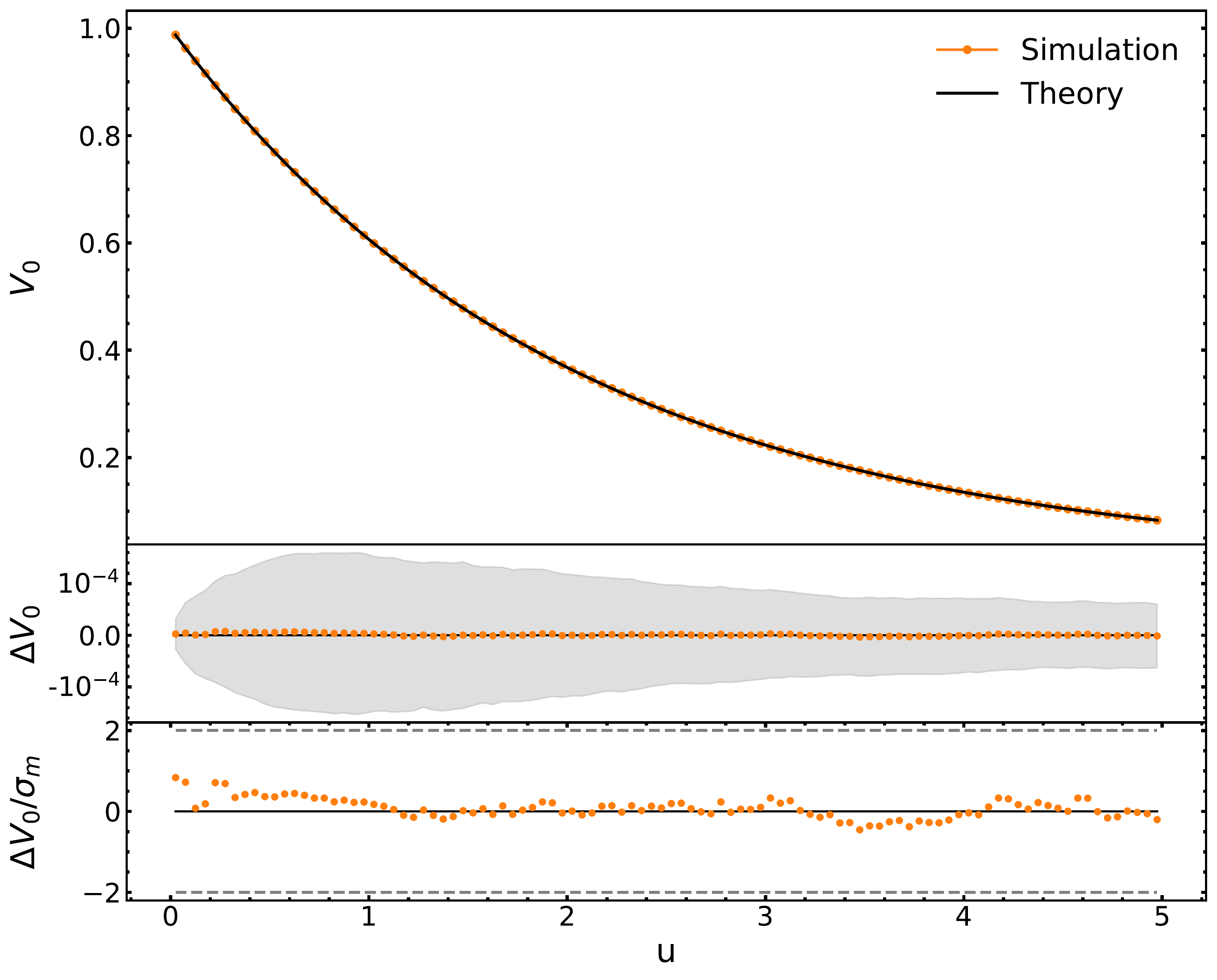} 
	
 \includegraphics[width=0.47\textwidth]{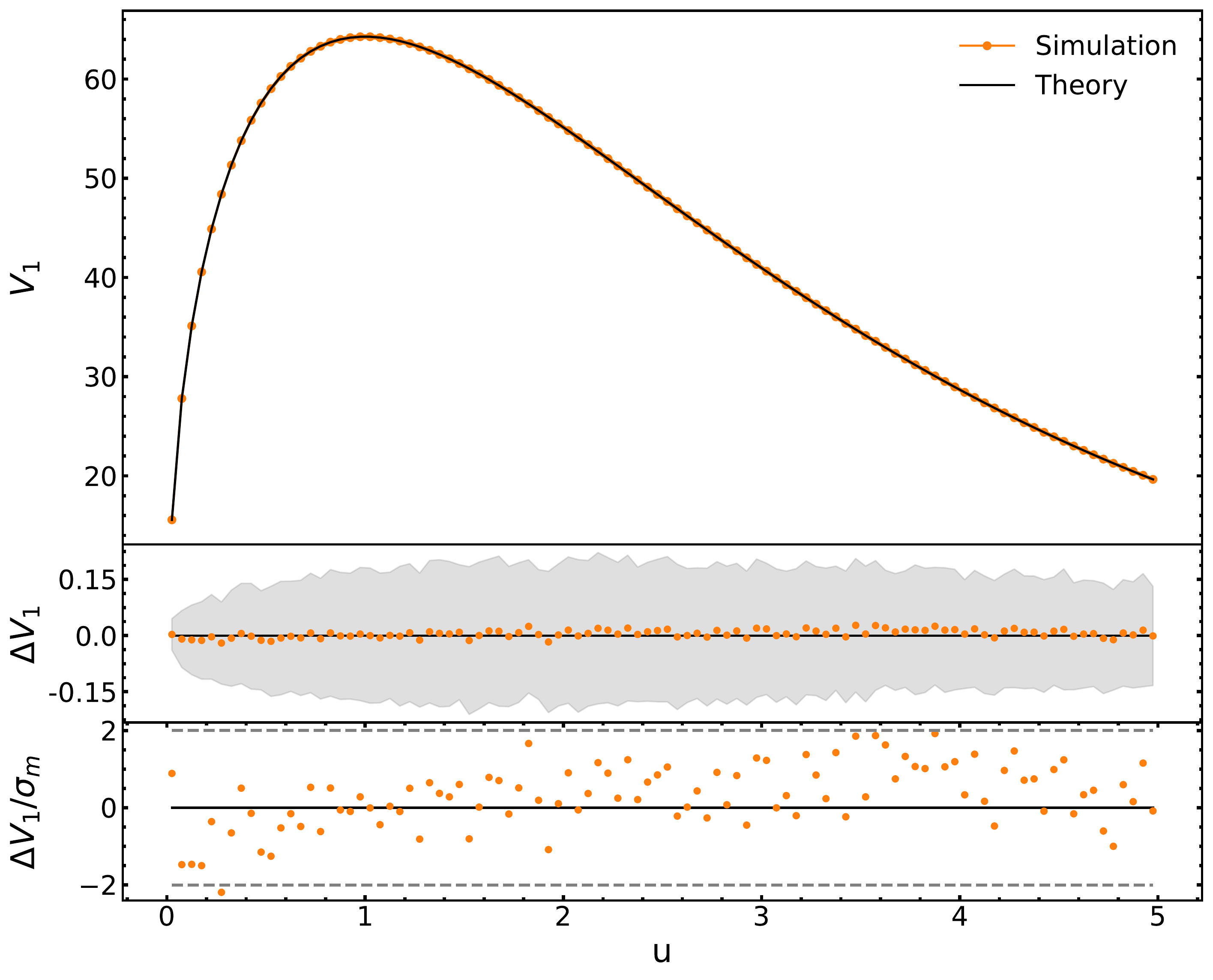} 
	
	\hspace{-0.19cm}\includegraphics[width=0.487\textwidth]{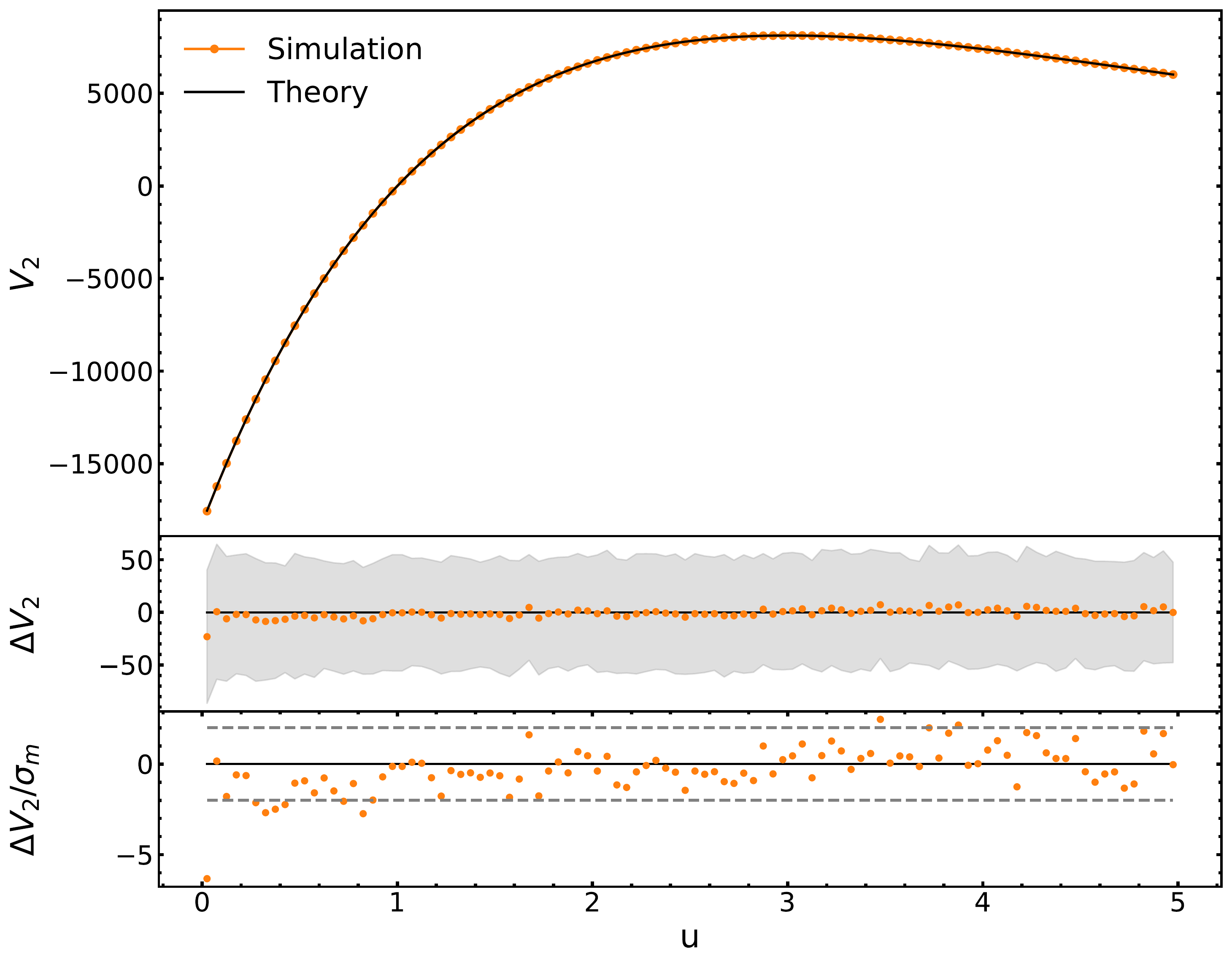} 
\caption{Average MFs of $P^2 = Q^2 + U^2$ for $N_{\textrm{sims}}=300$ CMB Gaussian isotropic simulations (orange dots) are compared with the theoretical predictions (solid black lines). From top to bottom: $V_0$, $V_1$, and $V_2$. The values of the MFs are shown in the top of each panel. The difference between average MFs from simulations and theory is compared with the dispersion among simulations ($1\sigma$, given by the grey contours) and the dispersion of the mean of simulations (computed as $\sigma/N_{\textrm{sims}}$) in the middle and bottom sections of each panel, respectively. See text for details.}
\label{fig:MFs_cmb_gauss}
\end{figure}
\subsection{Validation}
\label{ss:validation}

We verify the theoretical predictions for the MFs of $P^2$ introduced in \Cref{s:mf_P} by comparing them with the results of simulated CMB Gaussian isotropic maps. We simulate $300$ such maps as described in \Cref{ss:simsgau} and compute the MFs on them as explained in \Cref{s:code}. Maps are generated with \textit{Planck} best--fit angular power spectra, so $\mu$ can be exactly computed. We use this value to obtain the theoretical expectations of the MFs according to \cref{eqs:theo_exp}. We verify that the results hold for arbitrary angular power spectra, as long as there is no significant power in the scales corresponding to the pixel size.

The results of the MFs for these simulated maps can be seen in \Cref{fig:MFs_cmb_gauss}, together with the theoretical predictions. We find that the simulations are fully compatible with the theoretical expectations for all three MFs, with differences between the average of the simulations and the theory well inside the $1\sigma$ region, where $\sigma$ is the dispersion of the MFs among the $300$ simulations. 

Furthermore, in order to explore any possible systematic deviation, we compare the average residuals in the simulations with respect to the theoretical expectations considering the standard deviation of the mean ($\sigma_{\textrm{mean}} = \frac{\sigma_{\textrm{sims}}}{\sqrt{300}}$). This comparison is shown in the bottom section of each panel in \Cref{fig:MFs_cmb_gauss}. 
The three MFs are all perfectly compatible with theory, both in the case of individual simulations and of the average behaviour.

We also note the low statistical variation of these curves: the relative uncertainty $\sigma(V_i(u))/V_i(u)$ at every threshold $u$ is below $1\%$ and, in the case of $V_0$, it is below $1$ part in $1 000$. These values of the statistical standard deviation depend on the angular power spectrum of the studied map: we observe that the variance decreases when increasing the maximum multipole considered, $\ell_{max}$ (\textit{i.e.}, when increasing the resolution of the maps). This is also known to be the case for temperature maps, as quantitatively studied in \citet{fantaye2015}. Such low levels of fluctuations in the values of MFs make them suitable tools to detect deviations from Gaussianity and isotropy in real data.

These results provide a double validation: on the one hand, they verify the mathematical framework used to predict the expected values of the MFs of $P^2$ maps; on the other hand, they validate our implementation in the \texttt{Pynkowski} package. 
We have verified that masking the maps has a negligible effect on these results beyond slightly increasing the noise due to the smaller sky fraction. This robustness is expected because, unlike the angular power spectrum, MFs are purely local quantities. Thus, the effect of masking is completely negligible.

\subsection{Analysis of \textit{Planck} 2018 polarization maps}
\label{ss:planck_analysis}
Once both the theoretical predictions and our implementation to compute MFs have been validated, we can apply this formalism to the observed CMB polarization maps in order to assess the presence of any non--Gaussianity or deviation from statistical isotropy. 
In this work, as a proof of concept, we analyse the CMB maps reported by \textit{Planck} \citep{2020A&A...641A...4P}, produced with two different methods: SMICA (shown in this Section) and SEVEM (shown in \Cref{ap:sevem}); see \Cref{s:data} for details. We verify that our results are consistent on both complementary pipelines. 
\begin{figure}
	\centering
	\hspace{-0.12 cm}\includegraphics[width=0.478\textwidth]{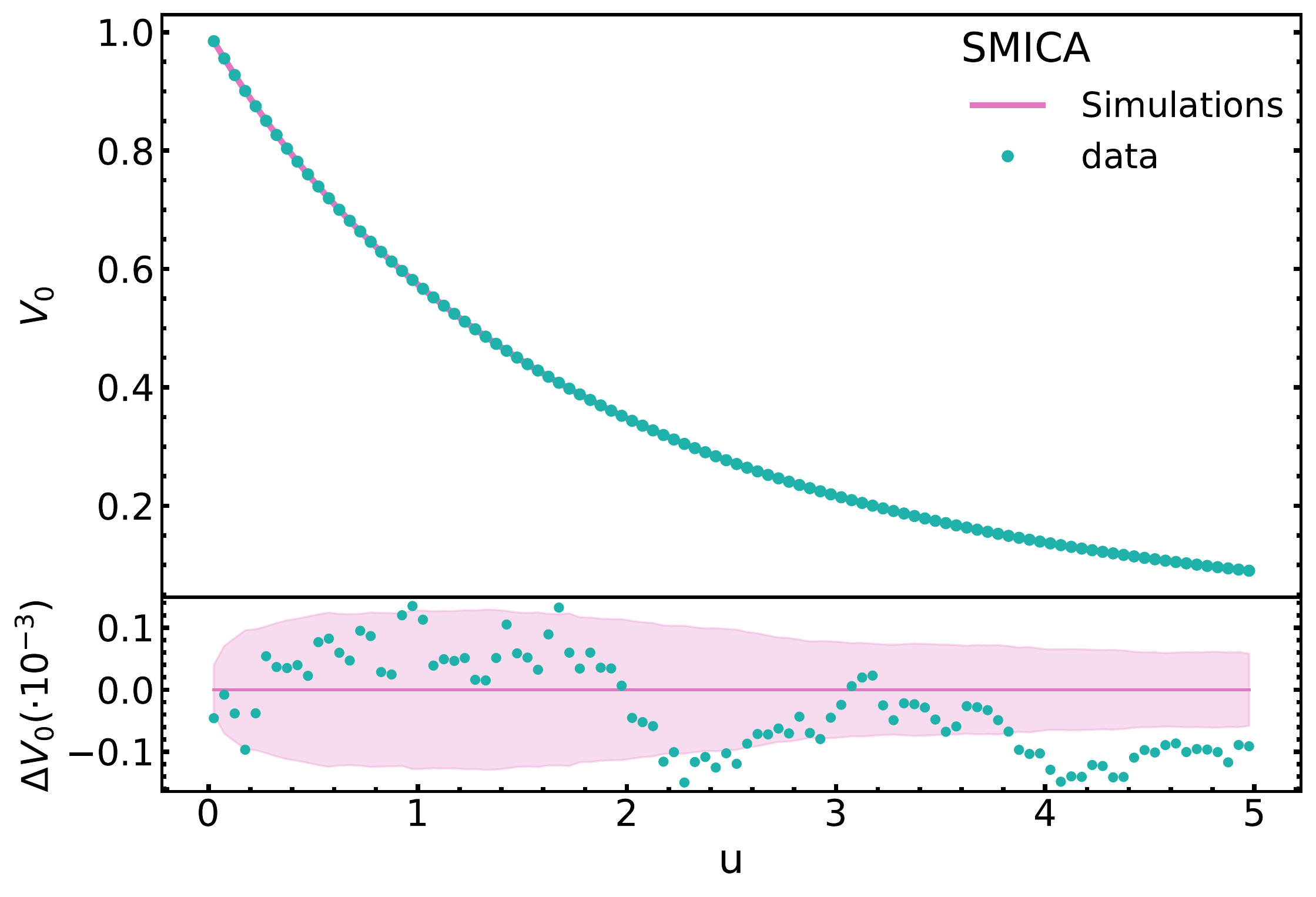} \\
	\hspace{0.1 cm}
     \includegraphics[width=0.455\textwidth]{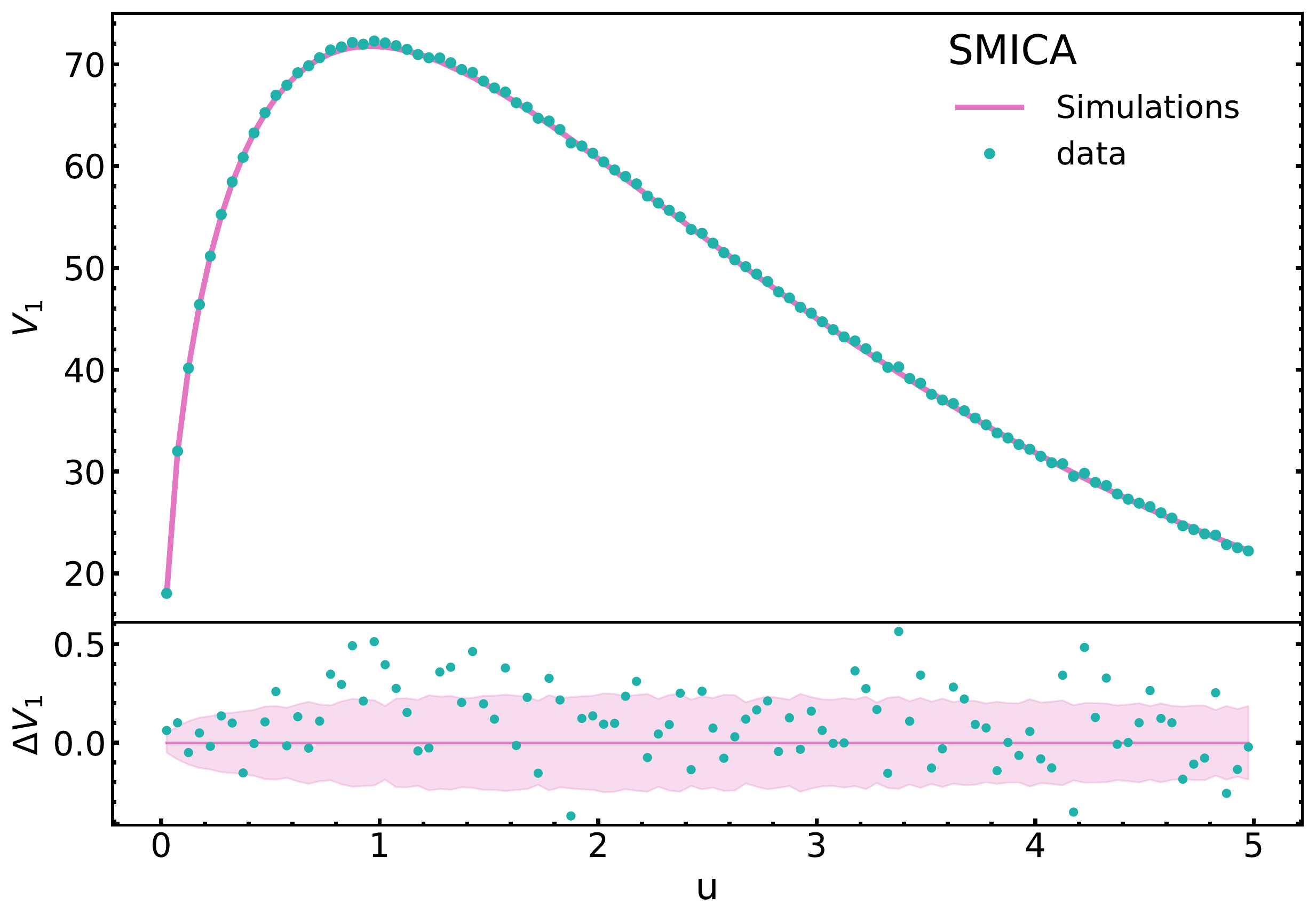}
	\includegraphics[width=0.47\textwidth]{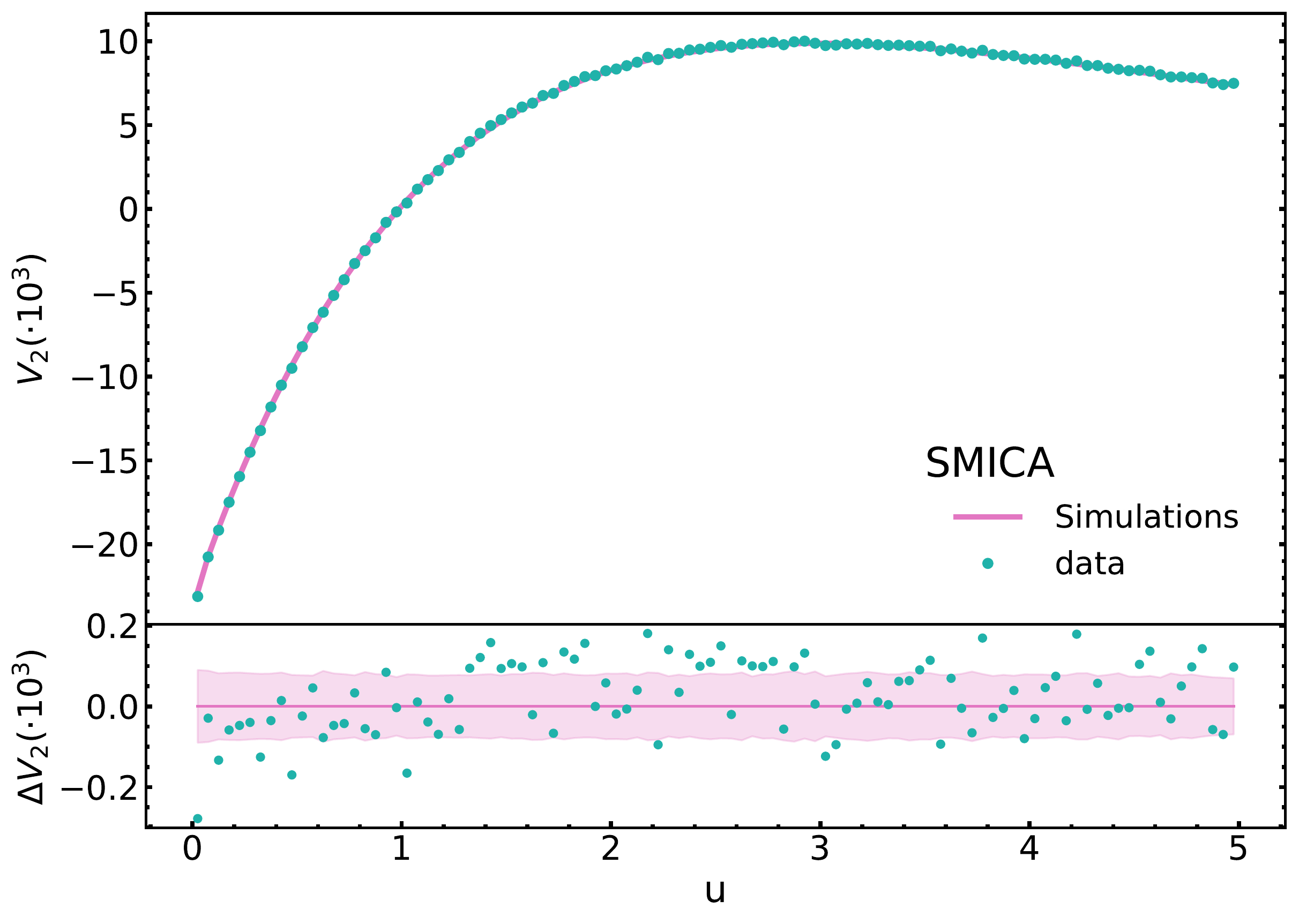} 
\caption{MFs of the \textit{Planck} SMICA released $P^2 = Q^2 + U^2$ maps (blue dots) are compared with the mean given by realistic end-to-end simulations (pink solid line). The first, second, and third MFs are shown (top to bottom), with the values (top section of each panel) and the residual with respect to the mean of simulations (bottom section of each panel). In all plots, the pink area represents the $\pm1\sigma$ region obtained in the simulations.}
\label{f:MFs_SMICA}
\end{figure}

\textit{Planck} polarization maps are affected by two known contaminating signals that would bias our MFs estimates if they are unaccounted for. The first is the residual emission from the Galactic foregrounds and point sources. In order to minimize their impact, we adopt the 2018 Component Separation Common polarization mask \citep{2020A&A...641A...4P}, shown in \Cref{fig:Planck mask}. The second contaminant is the residual noise whose statistical properties are not homogeneous across the sky due to the scanning strategy of the \textit{Planck} satellite \citep{2011A&A...536A...1P}, as can be seen in \Cref{fig:Planck maps}. This effect is consistently included in the end--to--end simulations provided by the collaboration, as explained in \Cref{ss:plancksims}. The impact of the anisotropic distribution of the statistical properties of noise is discussed in \Cref{ap:noise}.

In \Cref{f:MFs_SMICA}, we report the comparison between the MFs computed on the \textit{Planck} polarization modulus map and on end--to--end simulations, both produced by SMICA. We see for all MFs a remarkable agreement between them with all residuals within $\pm 2\sigma$. Only $V_0$ appears to be consistently below the average values in simulations at high thresholds, but this deviation is small and consistent with a statistical fluctuation.

We quantify the compatibility of all curves in terms of the $\chi^2$ statistic (as well as the probability--to--exceed and the deviation with respect to simulations in terms of $\Delta_{\chi}$), as explained in \Cref{ss:chi}. The results can be found in \Cref{t:chiscmb} and show that no deviation is statistically significant or consistently detected across different component separation algorithms. 

However, we can highlight some hints of deviation. The residuals of $V_0$ seem to present a pattern at high values of the threshold for both SMICA and SEVEM solutions. From the $\chi^2$ analysis, this trend is not significant and can be explained by the fact that the covariance matrix of $V_0$ presents significant off-diagonal terms, unlike the other two functionals. We note that this pattern is still present when adopting a more aggressive Galactic mask (with $f_{\textrm{sky}}=40\%$), thus disfavouring the hypothesis of a residual Galactic contamination origin.

So far, we have focused on \textit{Planck} $P^2$ maps which include the full range of multipoles. However, as already discussed in \Cref{ss:needlets}, applying an equivalent procedure in the needlet domain may provide a more detailed assessment separately at different angular scales. We apply a needlet transform on $Q \pm i U$ SMICA and SEVEM \textit{Planck} maps and corresponding end-to-end simulations and then construct needlet $P^2$ maps. This is done by adopting the configuration of needlet bands shown in \Cref{f:bands_nl}. We compute MFs of such obtained needlet maps and perform the same compatibility analysis of the full-range pixel-space case. The results for all needlet bands are shown in \Cref{t:chisPlanck_nl}, where we report the probability-to-exceed to ease the comparison with the analogous analysis performed for $E$ modes in \citet{2020A&A...641A...7P}.

As in the full-range pixel-space analysis we do not observe statistically significant deviations from the MFs of the simulated CMB and noise components at any scale. For SEVEM, we had to mask a $3\%$ of the sky along the Galactic plane before applying the needlet transform to avoid leaking of a residual foregrounds component outside the \textit{Planck} common mask in the first needlet scales. We used the publicly released \textit{Planck} \textit{GAL97} Galactic mask \footnote{\url{https://pla.esac.esa.int}}.

In conclusion, it is possible to state that, within the MFs sensitivity in $P^2$, no unexpected deviation from Gaussianity and statistical isotropy has been detected in the \textit{Planck} CMB $P^2$ maps. However, given the noise level in the \textit{Planck} polarization observations, such an analysis allows us to assess mainly the statistical properties of noise rather than those of the CMB signal. This is expected to change in the near future with upcoming high-sensitivity CMB experiments, as we shall show in the following section.

\begin{table}
\caption{Deviations of MFs computed on \textit{Planck} SMICA and SEVEM CMB $P^2$ maps with respect to those of the corresponding end-to-end simulations, in terms of normalised $\chi^2$, probability--to--exceed, and $\Delta_{\chi}$ of Eq. \ref{eq:delta_chi}.\label{t:chiscmb}}
\centering
\begin{tabular}{ll|ccc}
      && $\chi^2$  & $p_{\textrm{exc}}$ ($\%$)  & $\Delta_{\chi}$   \\ \hline
$V_0$ & SMICA &  1.02  &  43.7  & 0.04 \\
& SEVEM &  0.81  &  75.3  & -0.71  \\ \hline
$V_1$ & SMICA &  0.91 &  60.7  & -0.31  \\
& SEVEM &  1.36  &  8.3  & 1.48 \\ \hline
$V_2$ & SMICA &  1.22  & 21.0 & 0.72  \\
& SEVEM &  0.82  & 68.0 & -0.60
\end{tabular}
\end{table}

\begin{table*}
\caption{Deviations of MFs computed on \textit{Planck} SMICA and SEVEM CMB needlet $P^2$ maps with respect to those of the corresponding end-to-end simulations. They are reported in terms of the probability--to--exceed defined in \Cref{ss:chi}.\label{t:chisPlanck_nl}}
\centering
\begin{tabular}{ccccccccccccc}
& & & & & & & & & & $p_{\textrm{exc}}$ ($\%$) &
\end{tabular}\\
\begin{tabular}{|cc|cccccc|}
\hline 
      \multirow{2}{*}{Needlet scale} & \multirow{2}{*}{$\ell$ range}
      & \multicolumn{2}{c}{$V_0$} &
      \multicolumn{2}{c}{$V_1$} &
      \multicolumn{2}{c|}{$V_2$} \\
      & & SMICA & SEVEM & SMICA & SEVEM & SMICA & SEVEM \\
      \hline
 0 & $2-16$ &  46.7  & 43.0  &  15.3  &  12.3  & 11.3 & 4.3 \\
 1 & $8-32$ &  60.7  & 35.7  &  37.0  &  43.3  & 67.7 & 31.7 \\
 2 & $16-64$ &  47.0  & 75.3  &  83.0  &  87.3  & 79.7 & 86.3 \\
 3 & $32-128$ &  88.0  & 36.3  &  83.0  &  93.3  & 75.7 & 2.7 \\
 4 & $64-256$ &  64.3  & 90.7  &  82.3  &  53.7  & 31.3 & 38.0 \\
 5 & $128-512$ &  7.3  & 33.7  &  59.3  &  39.3  & 31.0 & 74.3 \\
 6 & $256-1024$ &  82.3  & 30.0  &  20.0  &  38.3  & 45.7 & 64.3 \\
 7 & $512-1024$ &  52.7  & 10.0  &  29.7  &  6.3  & 40.3 & 16.7 \\
 \hline
\end{tabular}
\end{table*}

\subsection{Forecasts for \textit{LiteBIRD} and SO}
\label{ss:fcasts}

The \textit{Planck} Collaboration provided cosmic-variance limited measurements of the CMB temperature anisotropies up to very high multipoles \citep{planck_spectra}. Conversely, in polarization, only modes at large angular scales have a modest signal-to-noise ratio. This is not the case for future surveys which will provide much more accurate polarization data. In this section, we present a forecast of the sensitivity of $P^2$ MFs to non-Gaussian anisotropic signals for two CMB experiments: \textit{LiteBIRD} and SO-SATs. One of the main challenges these experiments will need to face in their hunt for primordial $B$ modes will be an exquisite control of residual contamination by Galactic foregrounds, which is highly non-Gaussian and anisotropic. Therefore we test the ability of MFs applied to $P^2$ maps to detect such a signal with varying amplitude in simulated \textit{LiteBIRD} and SO-SATs data. The obtained results are compared with those of analogous \textit{Planck} simulated polarization maps. 
\begin{figure*}
	\centering
	\includegraphics[width=0.48\textwidth]{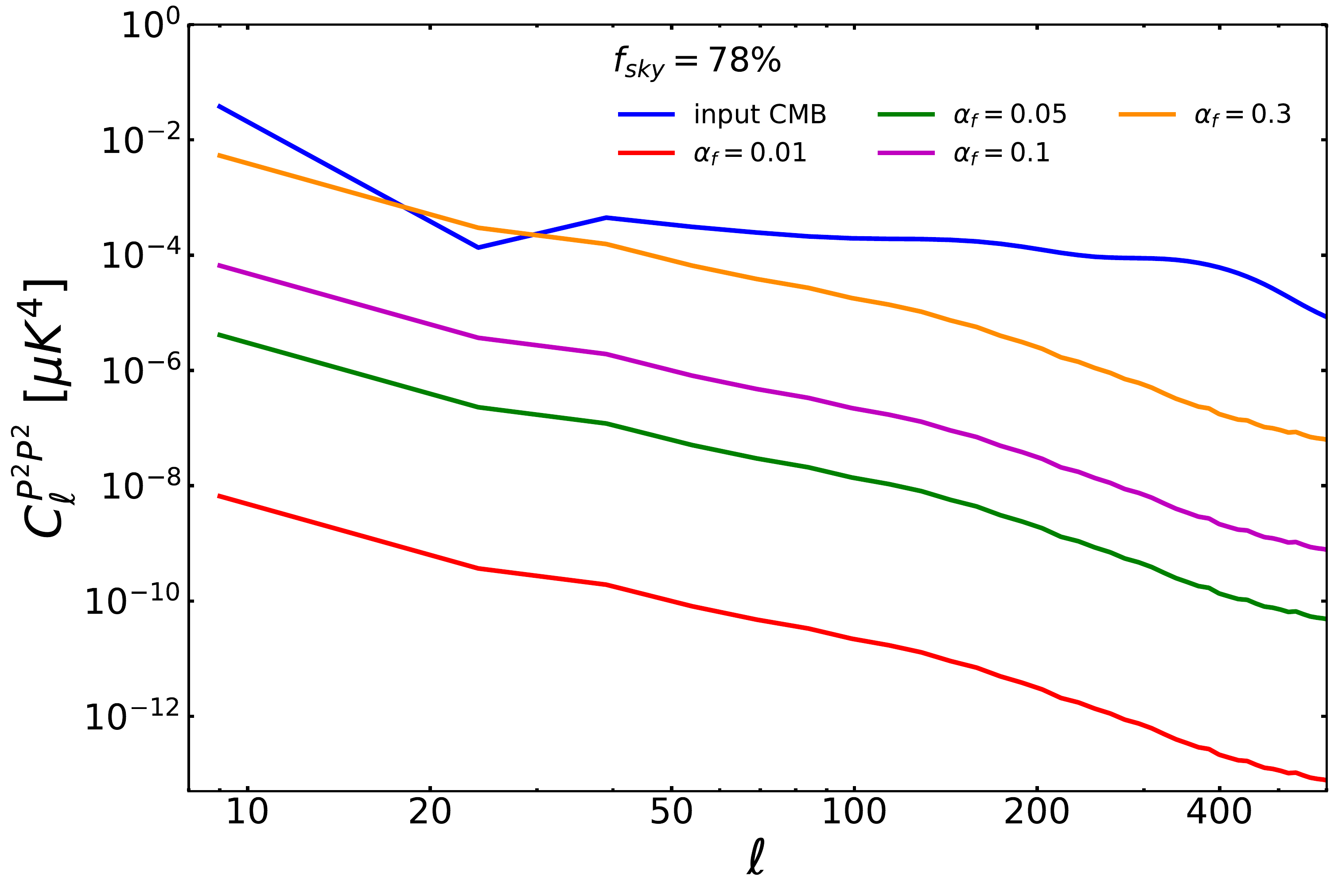}
	\includegraphics[width=0.48\textwidth]{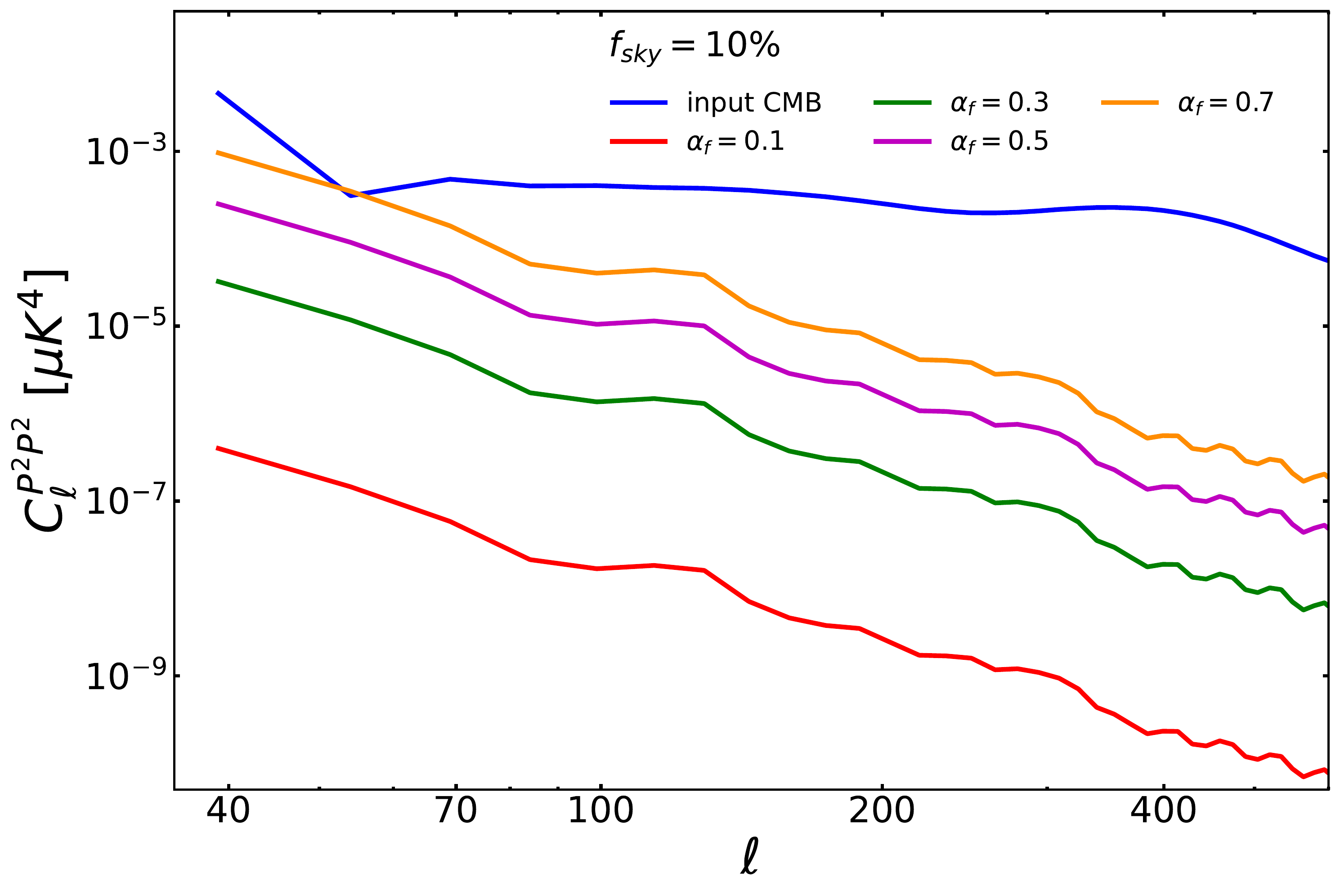}
\caption{Angular power spectra of $P^2$ maps computed from input $Q$ and $U$ maps employed to simulate the fiducial and test data sets for \textit{LiteBIRD}, SO-SATs, and \textit{Planck}. Blue solid lines represent input CMB, while other solid lines show power spectra of injected Galactic foreground emission with different values of the overall rescaling amplitude (readers can refer to the \Cref{ss:LB_SO_sims} for details). Left panel: Power spectra have been computed within the \textit{Planck} common polarization mask (see \Cref{fig:Planck mask}) and input maps have been smoothed with a Gaussian beam with FWHM$=37.8^{\prime}$. Right panel: Power spectra have been computed within the reduced SO-SATs patch (see \Cref{fig:patch_so}) and excluding the masked regions in the \textit{Planck} common polarization mask. In this case, input maps have been smoothed with a Gaussian beam with FWHM$=30.0^{\prime}$. The adopted binning scheme is $\Delta\ell =15$.}
\label{fig:Cl_P2_LB_SO}
\end{figure*}
\begin{table*}
\caption{Deviations of MFs computed on needlet $P^2$ maps of the test data set with respect to those of the Gaussian fiducial data set for \textit{LiteBIRD} and \textit{Planck}. MFs are computed within the \textit{Planck} common mask for polarization shown in \Cref{fig:Planck mask} ($f_{\textrm{sky}}\sim 78\%$). The deviations are expressed in terms of the average probability in percentage for a $\chi^2$ distribution with $N_{\textrm{dof}}=28$ (the adopted number of thresholds) to exceed the computed $\chi^2$s. Values lower than $10^{-6}$ are approximated with $0.0$. Values close to $50\%$ mean no detection of the injected non-Gaussian anisotropic signal, while values very close to $0\%$ mean detection at very high significance. Different magnitudes of the overall rescaling factor $\alpha_{\textrm{f}}$ of the residual foreground contamination are considered. 
\label{t:LB_vs_Planck_nl}}
\centering
\begin{tabular}{c}
\textbf{$\alpha_{\textrm{f}}=0.01$}
\end{tabular} \\
\begin{tabular}{|cc|cccccc|}
\hline 
      \multirow{2}{*}{Needlet scale} & \multirow{2}{*}{$\ell$ range}
      & \multicolumn{2}{c}{$V_0$} &
      \multicolumn{2}{c}{$V_1$} &
      \multicolumn{2}{c|}{$V_2$} \\
      & & \textit{Planck} & \textit{LiteBIRD} & \textit{Planck} & \textit{LiteBIRD} & \textit{Planck} & \textit{LiteBIRD} \\
      \hline
 0 & $2-16$ &  50.2 & 47.5 & 51.7 & 49.4 & 50.2 & 49.6 \\
 \hline
 1 & $8-32$ &  52.2 & 51.4 & 54.5 & 52.3 & 50.5 & 50.0 \\
 \hline
 2 & $16-64$ &  49.1 & 49.0 & 51.3 & 49.0 & 50.6 & 50.8
 \\
 \hline
 3 & $32-128$ &  49.9 & 50.2 & 51.9 & 50.0 & 50.1 & 49.4 \\
 \hline
 4 & $64-256$ &  49.6 & 50.8 & 50.8 & 50.0 & 49.9 & 48.8 \\
 \hline
 5 & $128-512$ &  49.1 & 50.4 & 49.1 & 50.2 & 49.5 & 51.9 \\
 \hline
\end{tabular}\vspace{0.2 cm}\\
\begin{tabular}{c}
\textbf{$\alpha_{\textrm{f}}=0.05$}
\end{tabular} \\
\begin{tabular}{|cc|cccccc|}
\hline 
      \multirow{2}{*}{Needlet scale} & \multirow{2}{*}{$\ell$ range}
      & \multicolumn{2}{c}{$V_0$} &
      \multicolumn{2}{c}{$V_1$} &
      \multicolumn{2}{c|}{$V_2$} \\
      & & \textit{Planck} & \textit{LiteBIRD} & \textit{Planck} & \textit{LiteBIRD} & \textit{Planck} & \textit{LiteBIRD} \\
      \hline
 0 & $2-16$ &  49.3 & 51.2 & 45.5 & 14.2 & 43.2 & 10.8 \\
 \hline
 1 & $8-32$ &  47.7 & 0.0  & 
54.2 & 0.01 & 53.7 & 0.4 \\
 \hline
 2 & $16-64$ &  49.4 & 45.7 & 52.9 & 48.1 & 54.8 & 50.3
 \\
 \hline
 3 & $32-128$ &  52.1 & 49.4 & 52.3 & 51.6 & 51.1 & 49.1 \\
 \hline
 4 & $64-256$ &  48.9 & 50.1 & 50.7 & 49.1 & 50.6 & 48.6 \\
 \hline
 5 & $128-512$ &  47.4 & 51.3 & 49.8 & 51.0 & 49.8 & 50.2 \\
 \hline
\end{tabular}\vspace{0.2 cm}\\
\begin{tabular}{c}
\textbf{$\alpha_{\textrm{f}}=0.1$}
\end{tabular} \\
\begin{tabular}{|cc|cccccc|}
\hline 
      \multirow{2}{*}{Needlet scale} & \multirow{2}{*}{$\ell$ range}
      & \multicolumn{2}{c}{$V_0$} &
      \multicolumn{2}{c}{$V_1$} &
      \multicolumn{2}{c|}{$V_2$} \\
      & & \textit{Planck} & \textit{LiteBIRD} & \textit{Planck} & \textit{LiteBIRD} & \textit{Planck} & \textit{LiteBIRD} \\
      \hline
 0 & $2-16$ &  49.2 & 53.5 & 20.8 & 0.07 & 20.4 & 0.6 \\
 \hline
 1 & $8-32$ &  42.7 & 0.0 & 
43.9 & 0.0 & 
49.9  & 0.0 \\
 \hline
 2 & $16-64$ &  45.3 & 37.9 & 53.9 & 23.7 & 52.3 & 23.5
 \\
 \hline
 3 & $32-128$ &  51.2 & 49.6 & 51.6 & 49.6 & 50.6 & 48.2 \\
 \hline
 4 & $64-256$ &  45.2 & 50.8 & 51.7 & 49.7 & 51.2 & 48.8 \\
 \hline
 5 & $128-512$ &  45.9 & 52.4 & 48.6 & 50.8 & 49.2 & 50.7 \\
 \hline
\end{tabular}\vspace{0.2 cm}\\
\begin{tabular}{c}
\textbf{$\alpha_{\textrm{f}}=0.3$}
\end{tabular} \\
\begin{tabular}{|cc|cccccc|}
\hline 
      \multirow{2}{*}{Needlet scale} & \multirow{2}{*}{$\ell$ range}
      & \multicolumn{2}{c}{$V_0$} &
      \multicolumn{2}{c}{$V_1$} &
      \multicolumn{2}{c|}{$V_2$} \\
      & & \textit{Planck} & \textit{LiteBIRD} & \textit{Planck} & \textit{LiteBIRD} & \textit{Planck} & \textit{LiteBIRD} \\
      \hline
 0 & $2-16$ &  52.1 &  29.3 & 
0.002 &  0. & 
0.7  &  0.001 \\
 \hline
 1 & $8-32$ &  0.0002 & 0.0 &
0.00004 & 0.0 &
1.6 & 0.0 \\
 \hline
 2 & $16-64$ &  26.0 & 0.0 &
10.0 & 0.0 &
11.9 & 0.0 \\
 \hline
 3 & $32-128$ &  52.5 & 43.2 & 
53.9 & 0.6 & 53.4 & 3.5 \\
 \hline
 4 & $64-256$ &  45.3 &  53.2 &
52.6 & 36.7 &
51.3 & 33.2 \\
 \hline
 5 & $128-512$ &  47.0 & 51.5 &
48.7 & 48.2 &
50.1 & 48.3
 \\
 \hline
\end{tabular}
\end{table*}

The \textit{LiteBIRD} and SO simulated data sets have been described in \Cref{ss:LB_SO_sims}. The analysis is performed in the needlet domain to assess the detection power of $P^2$ MFs at different angular scales. For each experiment, the MFs are computed on needlet maps of the fiducial and test data set. The fiducial data set includes only Gaussian CMB and noise components, while in the test data set we injected a non-Gaussian and anisotropic diffused Galactic emission with a rescaled overall amplitude (see \Cref{ss:LB_SO_sims}). Since such signal is particularly relevant on large and intermediate angular scales, we report results only for the first six needlet scales. As explained in \Cref{ss:chi}, we report for each case the average probability for a $\chi^2$ distribution with $N_{\textrm{dof}}=28$ (the adopted number of thresholds) to exceed the computed $\chi^2$s. 

\subsubsection*{\textit{LiteBIRD} vs \textit{Planck}}
\textit{LiteBIRD} will be the successor of the \textit{Planck} satellite with much improved sensitivity to CMB polarization. In this section we report the improvement in detecting any deviation from Gaussianity and statistical isotropy which is expected to be reached with \textit{LiteBIRD} when MFs are applied on maps of the polarization modulus. We compute MFs on $300$ $P^2$ simulations of the fiducial and test data set for both \textit{LiteBIRD} and \textit{Planck} within the \textit{Planck} $2018$ common polarization mask shown in \Cref{fig:Planck mask}. This analysis is performed in needlet domain. To match the HEALPix resolution of the needlet maps, the mask is downgraded accordingly, if needed, as described in \Cref{ss:needlets}. 

In this case, we consider the following values for $\alpha_{\textrm{f}}$: $\{0.01,\ 0.05,\ 0.1,\ 0.3\}$. Some of these amplitudes are in the range of the required subtraction of the Galactic emission to possibly detect CMB primordial $B$ modes. The angular power spectra of such rescaled non-Gaussian anisotropic foreground $P^2$ signal is compared with that of simulated CMB maps in the left panel of \Cref{fig:Cl_P2_LB_SO}. All angular power spectra are computed adopting the \textit{Planck} common mask of \Cref{fig:Planck mask}.

The obtained results for all needlet scales are shown in \Cref{t:LB_vs_Planck_nl}. For $\alpha_{\textrm{f}}=0.01$, the non-Gaussian signal is so low that $P^2$ MFs of the test data set are fully compatible with the Gaussian and isotropic hypotheses in both \textit{LiteBIRD} and \textit{Planck}. For a signal with $\alpha_{\textrm{f}}=0.05$, \textit{LiteBIRD} yields a highly significant detection at needlet scale $j=1$ with all MFs and a possible detection at the largest angular scales ($j=0$) in $V_1$ and $V_2$. No deviations, instead, are observed either at higher needlet scales or in \textit{Planck}. When $\alpha_{\textrm{f}}=0.1$, we find a clear detection for \textit{LiteBIRD} at $j=0$ in $V_1$ and $V_2$, at $j=1$ in all MFs, and hints for a detection at $j=2$ in $V_1$ and $V_2$, while, for \textit{Planck}, deviations only at $j=0$ in $V_1$ and $V_2$. Finally, with $\alpha_{\textrm{f}}=0.3$, for \textit{LiteBIRD} we have a detection at $j=0,\ 1,\ 2,$ and $3$, while for \textit{Planck} we can observe a detection at $j=0$ and $j=1$ and hints for a deviation at $j=2$. 

The reported results highlight the highly enhanced detection power of MFs if applied to \textit{LiteBIRD} $P^2$ data in comparison with \textit{Planck} at all the angular scales of interest. Therefore, with \textit{LiteBIRD}, we will have much better sensitivity to any possible residual contamination from foregrounds or other systematics and to cosmological effects which induce deviations from Gaussianity and statistical isotropy in the polarization data. We can note that, in general, $V_1$ and $V_2$ can better detect deviations than $V_0$. Moreover, for all the considered cases, unexpectedly we have more significant detections for $j=1$ rather than for $j=0$. This is motivated by the fact that the adopted \textit{Planck} common polarization mask better removes foreground signal on the largest angular scales. We have repeated the same analysis computing the MFs over the full sky and, indeed, the detection power at $j=0$ highly increases in this case. 

\subsubsection*{SO-SATs vs \textit{Planck}}
SO-SATs is one of the most promising near-future experiments for the observation of primordial $B$ modes from the ground. In this section, we compare the capability of detecting non-Gaussian anisotropic signals in the polarization modulus maps with MFs for SO-SATs and \textit{Planck}. We compute MFs on $300$ $P^2$ simulations of the fiducial and test data set for both SO-SATs and \textit{Planck} within the reduced SO-SATs patch shown in \Cref{fig:patch_so} excluding only the masked sources in the \textit{Planck} common polarization mask. This analysis is performed in needlet domain. Since SO-SATs will be sensitive to modes at $\ell > 30$, we consider only needlet scales with $j \ge 2$. On these smaller scales, the residual foreground emission, to be detectable, has to be more intense. Moreover, SO-SATs will observe only regions at high Galactic latitudes (see \Cref{fig:patch_so}), where the injected non-Gaussian anisotropic signal is expected to be fainter than that in the mask used for \textit{LiteBIRD} ($f_{\textrm{sky}}=78\%$). Therefore we adopt the following values for $\alpha_{\textrm{f}}$: $\{0.1,\ 0.3,\ 0.5,\ 0.7\}$. The angular power spectra of such rescaled non-Gaussian anisotropic foreground $P^2$ signal is compared with that of simulated CMB maps in the right panel of \Cref{fig:Cl_P2_LB_SO}. All angular power spectra are computed within the reduced SO-SATs sky patch of \Cref{fig:patch_so} and excluding the masked regions in the \textit{Planck} common polarization mask.

The obtained results are shown in \Cref{t:SO_vs_Planck_nl}. In the case of an overall rescaling factor $\alpha_{\textrm{f}}=0.1$, $P^2$ MFs are not sensitive to the injected non-Gaussian anisotropic signal in the SO-SATs patch neither for SO nor for \textit{Planck}. Instead, by computing $V_1$ and $V_2$ in SO-SATs data, we can detect deviations at $j=2$ for $\alpha_{\textrm{f}}=0.3$, at $j=2,\ 3$, and $4$ for $\alpha_{\textrm{f}}=0.5$ and $\alpha_{\textrm{f}}=0.7$, and hints for deviations at $j=3$ for $\alpha_{\textrm{f}}=0.3$. For the two largest considered rescaling amplitudes ($\alpha_{\textrm{f}}=0.5$ and $0.7$), also $V_1$ and $V_2$ of \textit{Planck} needlet $P^2$ test data set at $j=2$ significantly deviate from the fiducial Gaussian case. 

Also for SO-SATs the improved sensitivity will highly enhance the detection power of MFs in $P^2$ data with respect to the current \textit{Planck} maps within the SO observed patch. As in the comparison between \textit{LiteBIRD} and \textit{Planck}, the first MF $V_0$ results less sensitive to non-Gaussian anisotropic signals than $V_1$ and $V_2$. In this analysis we assumed a pessimistic 1/f noise component for SO-SATs. We thus expect even a higher detection power in the first needlet bands in case of improved instrumental properties. 

\begin{table*}
\caption{Deviations of MFs computed on needlet $P^2$ maps of the test data set with respect to those of the Gaussian fiducial data set for SO-SATs and \textit{Planck}. MFs are computed within the reduced SO-SATs patch shown in \Cref{fig:patch_so} and excluding the masked regions of the \textit{Planck} polarization common mask ($f_{\textrm{sky}}\sim 10\%$). The deviations are expressed in terms of the average probability in percentage for a $\chi^2$ distribution with $N_{\textrm{dof}}=28$ (the adopted number of thresholds) to exceed the computed $\chi^2$s. Values lower than $10^{-6}$ are approximated with $0.0$. Values close to $50\%$ mean no detection of the injected non-Gaussian anisotropic signal, while values very close to $0\%$ mean detection at very high significance. Different magnitudes of the overall amplitude $\alpha_{\textrm{f}}$ of the residual foreground contamination are considered. 
\label{t:SO_vs_Planck_nl}}
\centering
\begin{tabular}{c}
\textbf{$\alpha_{\textrm{f}}=0.1$}
\end{tabular} \\
\begin{tabular}{|cc|cccccc|}
\hline 
      \multirow{2}{*}{Needlet scale} & \multirow{2}{*}{$\ell$ range}
      & \multicolumn{2}{c}{$V_0$} &
      \multicolumn{2}{c}{$V_1$} &
      \multicolumn{2}{c|}{$V_2$} \\
      & & \textit{Planck} & SO-SATs & \textit{Planck} & SO-SATs & \textit{Planck} & SO-SATs \\
      \hline
  2 & $30-64$ &  47.9  & 53.1  &  54.8  &  33.2  & 56.0 & 39.0 \\
 \hline
 3 & $32-128$ &  50.6 & 51.4 & 52.6 & 51.3 & 49.7 & 50.3 \\
 \hline
 4 & $64-256$ &  48.9 & 46.4 & 50.1 & 49.6 & 50.5 & 49.5 \\
 \hline
 5 & $128-512$ &  49.4 & 48.6 & 49.2 & 50.4 & 49.2 & 48.4 \\
 \hline
\end{tabular}\vspace{0.2 cm}\\
\begin{tabular}{c}
\textbf{$\alpha_{\textrm{f}}=0.3$}
\end{tabular} \\
\begin{tabular}{|cc|cccccc|}
\hline 
      \multirow{2}{*}{Needlet scale} & \multirow{2}{*}{$\ell$ range}
      & \multicolumn{2}{c}{$V_0$} &
      \multicolumn{2}{c}{$V_1$} &
      \multicolumn{2}{c|}{$V_2$} \\
      & & \textit{Planck} & SO-SATs & \textit{Planck} & SO-SATs & \textit{Planck} & SO-SATs \\
      \hline
  2 & $30-64$ &  50.3 & 54.2 & 35.3 & 0. &
26.2 & 0.003 \\
 \hline
 3 & $32-128$ &  52.8 & 49.7 & 56.1 & 18.0 & 54.0 & 31.9 \\
 \hline
 4 & $64-256$ &  52.6 & 49.6 & 49.7 & 48.6 & 49.6 & 49.2 \\
 \hline
 5 & $128-512$ &  48.5 & 50.0 & 49.8 & 48.6 & 
49.0 & 49.3 \\
 \hline
\end{tabular}\vspace{0.2 cm}\\
\begin{tabular}{c}
\textbf{$\alpha_{\textrm{f}}=0.5$}
\end{tabular} \\
\begin{tabular}{|cc|cccccc|}
\hline 
      \multirow{2}{*}{Needlet scale} & \multirow{2}{*}{$\ell$ range}
      & \multicolumn{2}{c}{$V_0$} &
      \multicolumn{2}{c}{$V_1$} &
      \multicolumn{2}{c|}{$V_2$} \\
      & & \textit{Planck} & SO-SATs & \textit{Planck} & SO-SATs & \textit{Planck} & SO-SATs \\
      \hline
  2 & $30-64$ &  51.6 & 47.5 & 0.04 & 0.0 &
0.4 & 0.0 \\
 \hline
 3 & $32-128$ &  50.3 & 51.4 & 54.0 & 0.0 & 50.7 & 0.3 \\
 \hline
 4 & $64-256$ &  50.7 & 49.7 & 51.5 & 0.0 & 52.3 & 4.2 \\
 \hline
 5 & $128-512$ &  47.8 & 48.8 & 49.0 & 47.7 & 48.9 & 48.0 \\
 \hline
\end{tabular}\vspace{0.2 cm}\\
\begin{tabular}{c}
\textbf{$\alpha_{\textrm{f}}=0.7$}
\end{tabular} \\
\begin{tabular}{|cc|cccccc|}
\hline 
      \multirow{2}{*}{Needlet scale} & \multirow{2}{*}{$\ell$ range}
      & \multicolumn{2}{c}{$V_0$} &
      \multicolumn{2}{c}{$V_1$} &
      \multicolumn{2}{c|}{$V_2$} \\
      & & \textit{Planck} & SO-SATs & \textit{Planck} & SO-SATs & \textit{Planck} & SO-SATs \\
      \hline
  2 & $30-64$ &  49.3 & 26.8 & 0.0 & 0.0
& 0.0 & 0.0 \\
 \hline
 3 & $32-128$ &  50.1 & 47.1 & 26.0 & 0.0 & 
26.0 & 0.0 \\
 \hline
 4 & $64-256$ &  50.6 & 51.0 &
52.1 & 0.0 &
52.1 & 0.0
 \\
 \hline
 5 & $128-512$ &  46.7 & 47.6 & 48.8 & 46.3 & 48.8 & 46.8 \\
 \hline
\end{tabular}
\end{table*}

\section{Conclusions}
\label{s:concl}
	
MFs have been widely applied to the analysis of CMB data \citep{schmalzing1998,2020A&A...641A...7P}, but their application to polarization has been limited to date. In this work, we introduce a generalised framework to compute the MFs on maps of the squared modulus of polarization ($P^2 =Q^2+U^2$). MFs are able to provide information complementary to that obtained from angular power spectra and, in particular, it is sensitive to deviations from the hypothesis of a Gaussian isotropic field. Therefore, they are expected to be able to blindly detect: i) early- and late-time cosmological effects (\textit{e.g.} primordial non-Gaussianity or CMB lensing), ii) significant residual contamination by foregrounds and instrumental systematics.

The statistical analysis of the polarized intensity is complementary to that performed on E or B modes, as the CMB $P$ map reflects the local morphological properties of the polarization field. Therefore, we expect the introduced formalism to better capture those deviations which primarily affect the intensity rather than the direction of the polarization field. Furthermore, it allows us to easily study polarization data in the presence of any mask without suffering from any E--B and B--E leakage contamination \citep{2001PhRvD..65b3505L,2003PhRvD..67b3501B}. 
An alternative formalism to study the spin--$2$ CMB polarization field has recently been introduced in \citet{2023carron}.

Our main contributions are the following:
\begin{itemize}
    \item We introduce a general formalism to predict the value of MFs for $P^2$ in the case of a Gaussian and isotropic field, justifying how the effects of the spin are negligible in the cosmological case.
    \item We implement and release \texttt{Pynkowski}, a public Python package to compute MFs on HEALPix maps of any scalar field, such as CMB $T$ and $P^2$.
    \item We validate theoretical predictions for MFs from $P^2$ on Gaussian isotropic CMB simulations.
    \item We apply the formalism to \textit{Planck} polarization modulus maps (produced by SMICA and SEVEM) both in pixel-space and in needlet domain. We find compatibility (with less than $1.5$$\sigma$ deviation in all cases)  between their MFs and those computed on realistic end--to--end simulations, which take into account the anisotropic noise residuals due to the scanning strategy of the satellite. We can conclude that we do not observe any hint for deviations from Gaussianity or statistical isotropy in \textit{Planck} CMB polarization maps at any angular scale. However, given the noise level in the \textit{Planck} $P^2$ maps, such an analysis allows us to assess mainly the statistical properties of noise. 
\end{itemize}

In the coming years, we expect new high--quality CMB polarization observations from ongoing or planned experiments such as SPT \citep{2020PhRvD.101l2003S}, LSPE \citep{2021JCAP...08..008A}, Simons Observatory \citep{2019JCAP...02..056A}, CMB-S4 \citep{2022arXiv220308024A}, and \textit{LiteBIRD} \citep{hazumi2019}. MFs can play a key role in detecting unexpected non-Gaussian components or departures from statistical isotropy in these polarization maps. To further assess this point, we forecast the detection power of $P^2$ MFs for two future and complementary CMB experiments, the \textit{LiteBIRD} satellite and the ground-based SO-SATs, which will target the observation of primordial CMB polarization $B$ modes. We show that, when applied to simulated maps of the polarization modulus of \textit{LiteBIRD} and SO, MFs are much more sensitive to the presence of a non-Gaussian anisotropic signal than \textit{Planck}. We have shown this by using a residual Galactic foreground contamination model with varying amplitudes as an example of non-Gaussian anisotropic component.

Furthermore, MFs might be especially relevant in light of the large--scale anomalies that are being detected in multiple observables \citep{2013JCAP...07..047G,2016CQGra..33r4001S,2016PhRvD..93b3524A}. The introduced formalism may also be useful to study other complex spin fields defined on the sphere, such as the convergence ($\kappa$) or the gravitational shear ($\sigma$) maps associated to weak lensing. Lensing maps, indeed, are found to be highly non--Gaussian, as they are produced by the non-linear distribution of matter in the Universe \citep{zuercher2021}. 

Finally, MFs can be employed to study polarized Galactic and extragalactic foregrounds, as they strongly deviate from a Gaussian distribution and are (in the Galactic case) highly anisotropic. MFs can complement other higher--order statistics that are already used to analyse foregrounds, such as peak statistics \citep{carronduque2019} or neural networks \citep{farsian2020}. As an example of this synergy, in \citet{krachmalnicoff2020} they train a neural network to generate a high--resolution dust foreground map with a given shape of their MFs, as dust emission is highly non--Gaussian \citep{2018JCAP...11..047J, 2015JCAP...11..019B}.

The \texttt{Pynkowski} Python package we have developed to study MFs is fully documented and can be found in \url{https://github.com/javicarron/pynkowski}, along with examples of its use.
	
\section*{Acknowledgements}
MM and NV acknowledge support by ASI/COSMOS grant n. 2016-24-H.0 and grant n. 2020-9-HH.0. JCD and MM ackowledge support from ASI/INFN grant n. 2021-43-HH.0. Part of this work was also supported by the InDark INFN project. DM acknowledges support from the MIUR Excellence Project awarded to the Department of Mathematics, Università di Roma Tor Vergata, CUP E83C18000100006. DM is also grateful to the Department of Excellence Programme MatModTov for support.

 \section*{Data Availability}
The \textit{Planck} data used in this article are available in the \textit{Planck} Legacy Archive, at \url{https://pla.esac.esa.int/pla/}. The results produced in this article will be shared on reasonable request to the corresponding author. The software is available at \url{https://github.com/javicarron/pynkowski}.

\bibliographystyle{mnras}
\bibliography{final}

\clearpage
\appendix

\section{Robustness to CMB component separation algorithm}
\label{ap:sevem}

\begin{figure}
	\centering
	\hspace{-0.12 cm}\includegraphics[width=0.478\textwidth]{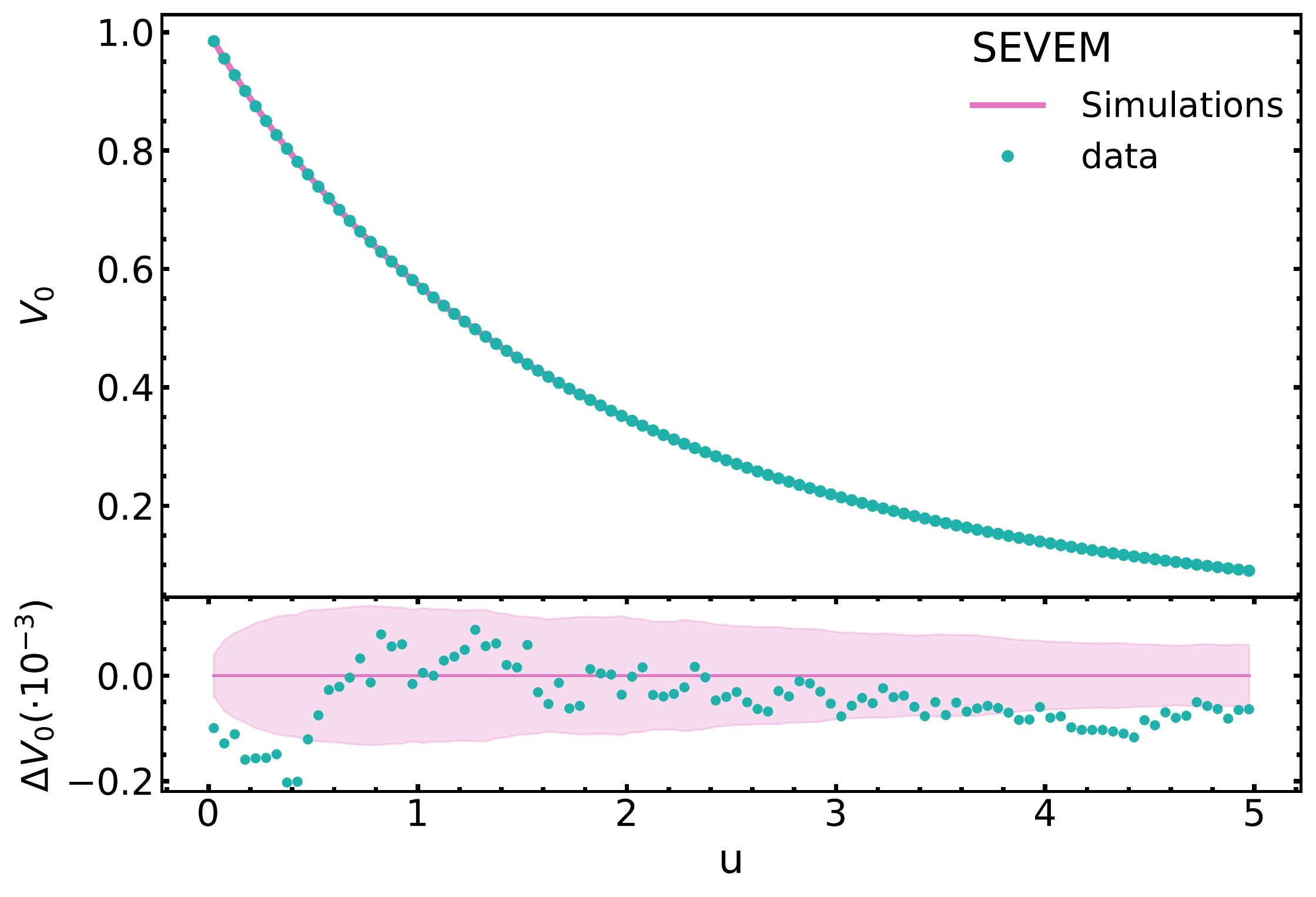} \\
     \includegraphics[width=0.47\textwidth]{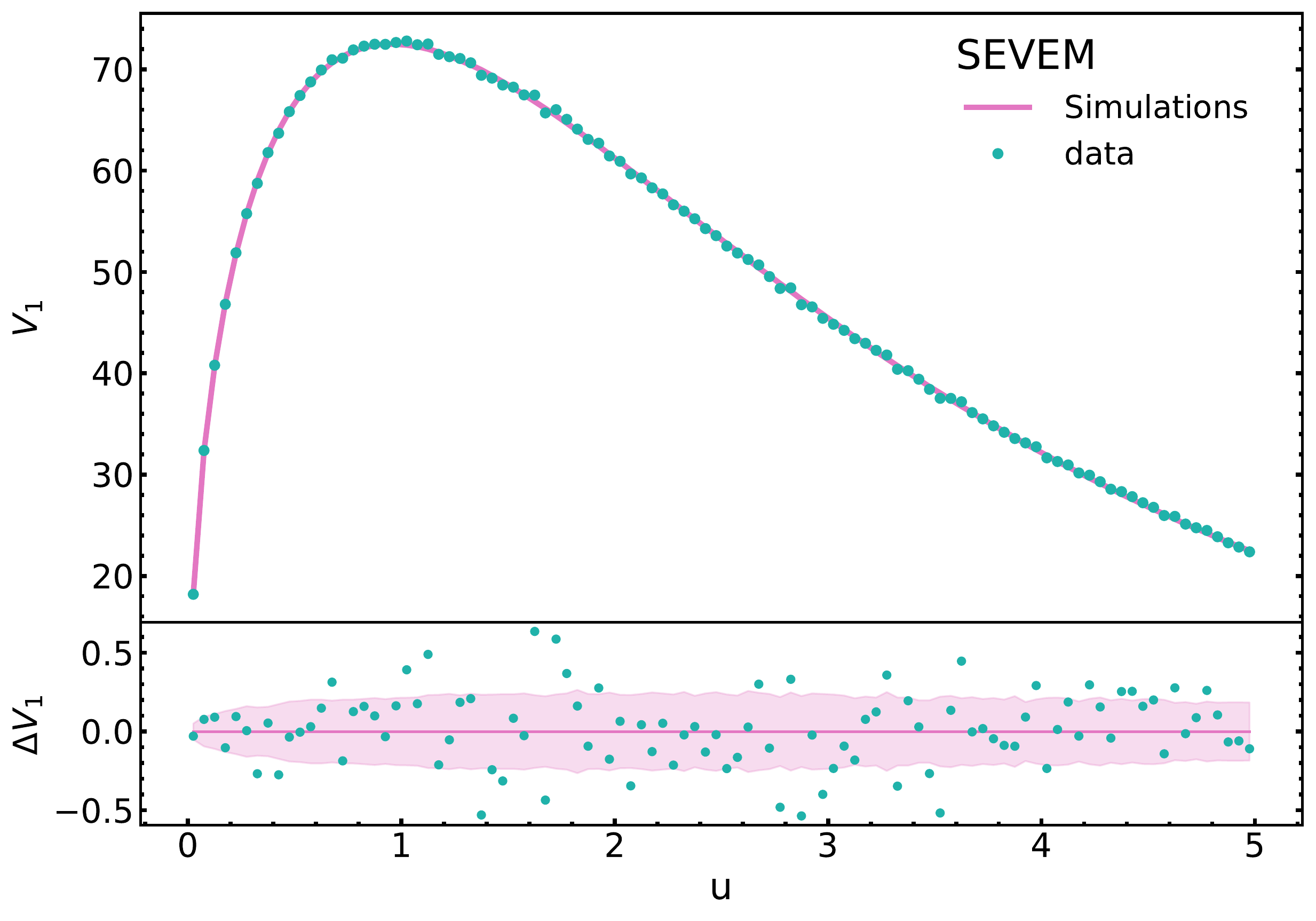}
	\includegraphics[width=0.47\textwidth]{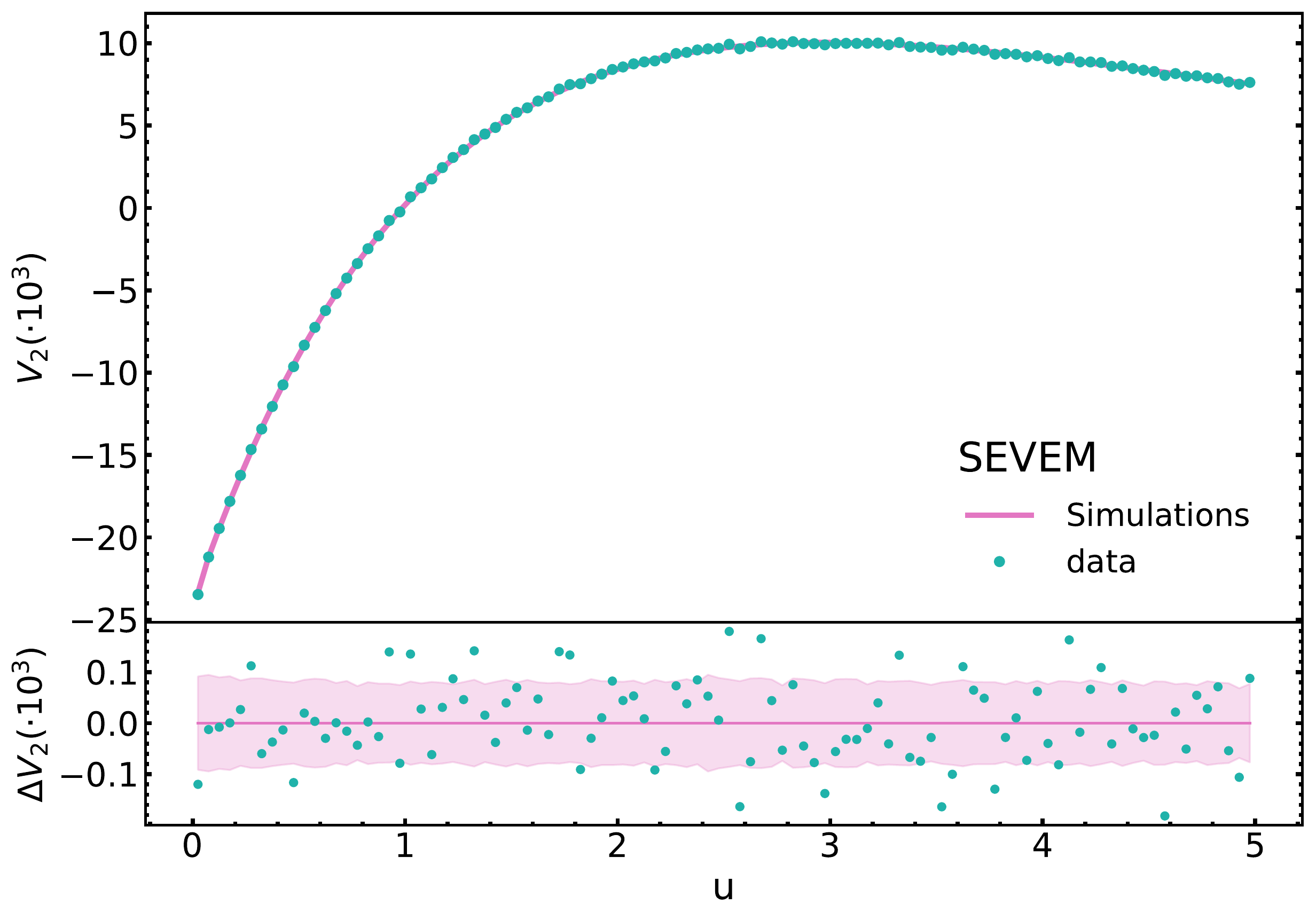} 
\caption{MFs of the \textit{Planck} SEVEM released $P^2 = Q^2 + U^2$ maps (blue dots) are compared with the mean given by realistic end-to-end simulations (pink solid line). The first, second, and third MFs are shown (top to bottom), with the values (top section of each panel) and the residual with respect to the mean of simulations (bottom section of each panel). In all plots, the pink area represents the $\pm1\sigma$ region obtained in the simulations.}
\label{f:MFs_SEVEM}
\end{figure}

\begin{figure}
	\centering
	\hspace{-0.08 cm}\includegraphics[width=0.48\textwidth]{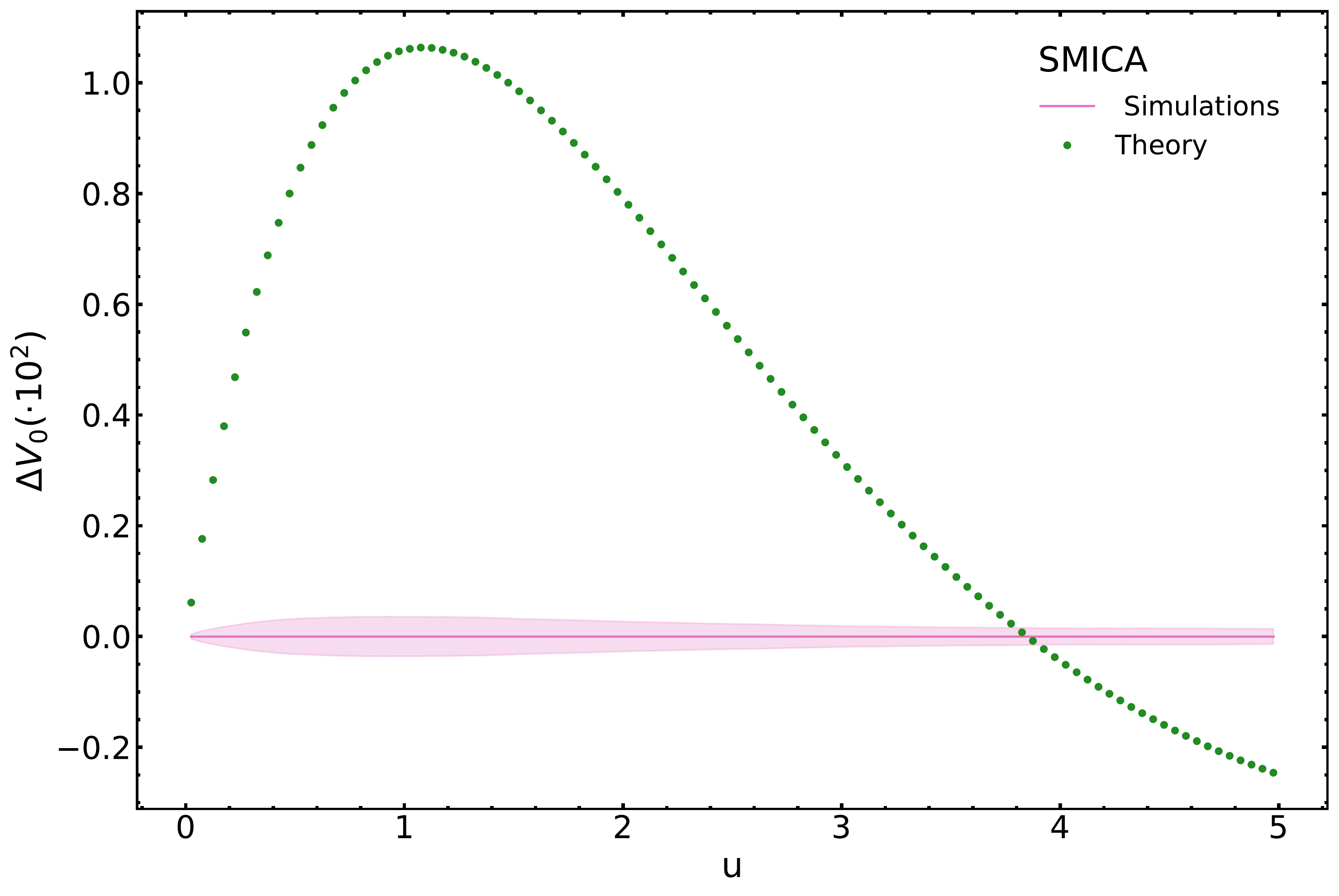}\\
	\includegraphics[width=0.48\textwidth]{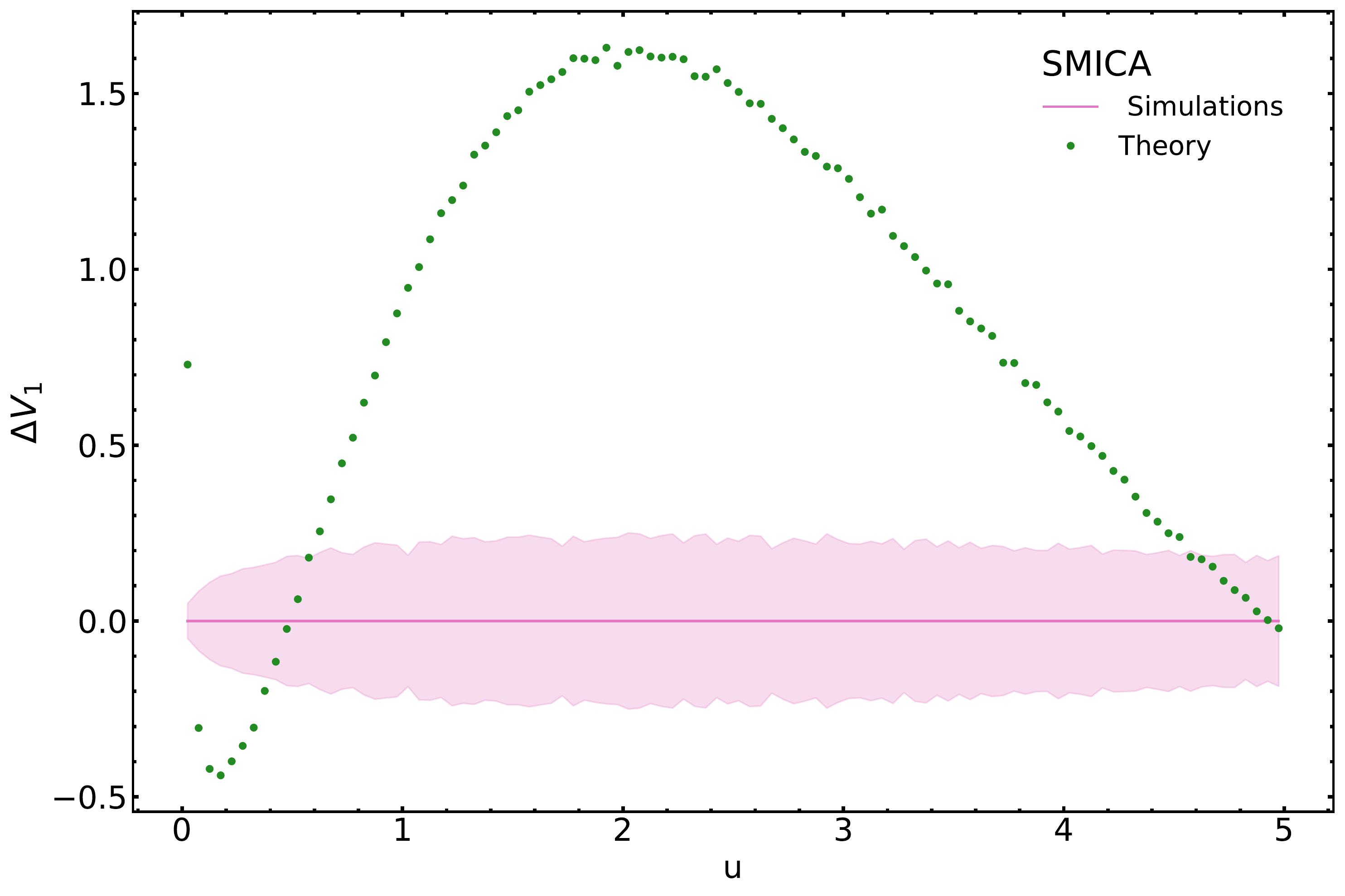}\\
    \hspace{0.0489 cm}
    \includegraphics[width=0.48\textwidth]{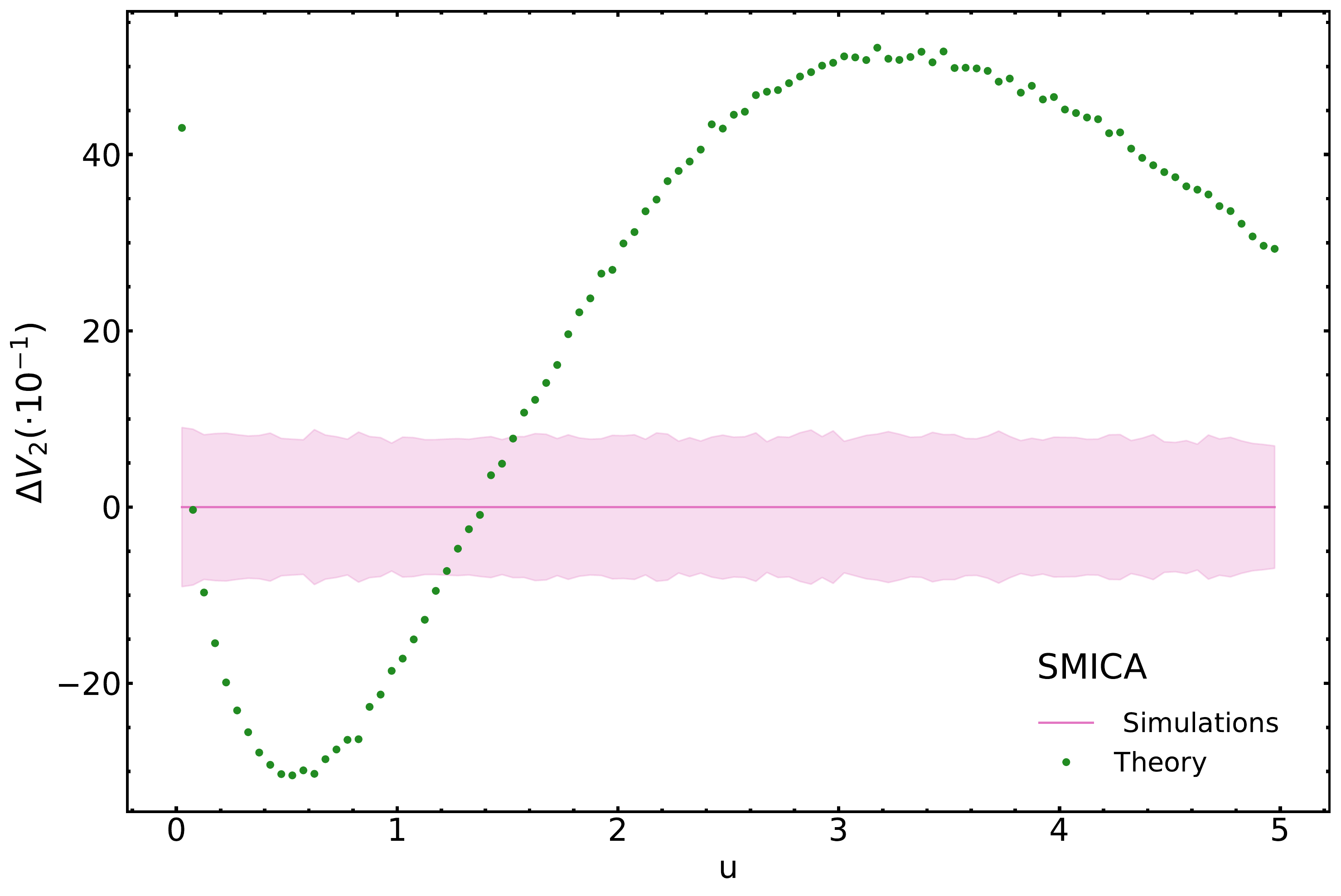} 
 
\caption{Deviations of theoretical MFs (green dots) in the Gaussianity and statistical isotropy hypotheses with respect to those computed on the \textit{Planck} SMICA $P^2 = Q^2 + U^2$ end-to-end simulations (pink). The first, second, and third MFs are shown (top to bottom). In all plots, the pink area represents the $\pm1\sigma$ region obtained in the simulations.}
\label{f:MFs_SMICA_gauss}
\end{figure}

In \Cref{ss:planck_analysis} we showed the results when MFs are computed on CMB \textit{Planck} polarization maps and realistic end--to--end simulations, both produced by the SMICA cleaning algorithm, assuming primordial Gaussianity and statistical isotropy, obtaining a complete compatibility between them.
We perform the same analysis with the SEVEM reconstructed maps in order to verify that the results are robust to the specific method used to extract the CMB signal.  

In \Cref{f:MFs_SEVEM} we show the comparison between MFs of the \textit{Planck} polarization modulus map and the ones of realistic simulations obtained with the SEVEM pipeline. As with SMICA, we can observe full compatibility between them with deviations always within $\pm 2\sigma$ where $\sigma$ represents the dispersion among the simulations. Therefore, we find no indication of primordial non--Gaussianity or deviations from statistical isotropy of the observed CMB polarization modulus.

\section{Impact of anisotropic noise}
\label{ap:noise}
\textit{Planck} CMB polarization maps are known to be affected by noise residuals that are significantly anisotropic due to the mission scanning strategy  \citep{2011A&A...536A...1P}, as it can be seen in \Cref{fig:Planck maps}. Some portions of the sky are better observed; this reduces the noise level and, consequently, the variance of the map in these regions. The anisotropic noise pattern is clearly visible in the SMICA and SEVEM $P^2$ maps. 
Hence, it needs to be properly modelled and accounted for when analysing the data looking for non-Gaussianity and departure from isotropy of primordial origin. The impact of anisotropic noise has been studied for \textit{Planck} CMB temperature by \citet{ducout2012}. To stress this point, let us study the impact of anisotropic noise contamination by comparing the MFs of SMICA and SEVEM realistic simulations with the theoretical expected values under the Gaussianity and isotropy hypotheses computed considering the average value of $\mu$ among all simulations  and using \cref{eqs:theo_exp}.

We follow an analogous procedure to the one in \Cref{ss:planck_analysis}. The deviations of theoretical expectations with respect to MFs computed on SMICA end--to--end simulations are reported in \Cref{f:MFs_SMICA_gauss}. The deviations are strongly significant at a level of $20-45\sigma$, detectable in all MFs and both component separation algorithms. It is remarkable that this strong deviation can be reproduced exactly by realistic noise simulations (compare Figs. \ref{f:MFs_SMICA} and \ref{f:MFs_SEVEM}, where noise+CMB simulations are perfectly compatible with observations).
This result highlights the potential of the application of MFs to future data, which will not be dominated by anisotropic noise as strongly as \textit{Planck} maps. It also underlines the importance of calibration to realistic simulations, especially in polarization data. Any possible future claim of detection of primordial non--Gaussianity or deviations from statistical isotropy of polarized CMB signal will have to be calibrated not only to realistic noise, but also to possible realistic residuals of galactic and extragalactic foregrounds.

\label{lastpage}	
\end{document}